\documentclass[11pt,english]{article}
\usepackage[a4paper,margin=1in]{geometry}
\usepackage{gensymb}
\usepackage{subcaption}
\usepackage{amssymb}
\usepackage{amsmath}
\usepackage{amsthm}
\usepackage{amsfonts}
\usepackage{verbatim}
\usepackage{graphicx}
\usepackage{tikz}
\usetikzlibrary{calc, backgrounds,arrows,patterns}

\theoremstyle{plain}
\newtheorem{theorem}{Theorem}[section]
\newtheorem{lemma}[theorem]{Lemma}

\theoremstyle{definition}

\theoremstyle{remark}

\usepackage{mathtools}

\DeclarePairedDelimiter\floor{\lfloor}{\rfloor}
\usepackage[square,sort,comma,numbers]{natbib}
\begin{document}

\title{\Large 
A 3D Advancing-Front Delaunay Mesh Refinement Algorithm
\thanks{The work of the author was supported in part by the NIH/NIGMS Center
for Integrative Biomedical Computing grant 2P41 RR0112553-12.  The author would
also like to thank Ms. Christine Pickett for proofreading a draft of
the paper and suggesting numerous changes. 
}}
\author{Shankar P. Sastry\\Google LLC, Sunnyvale, CA 94089, U.S.A.\\{\tt sastry@sci.utah.edu}}

\date{}

\maketitle
\begin{abstract}
I present a 3D advancing-front mesh refinement algorithm that generates a constrained Delaunay mesh for any piecewise linear complex (PLC) and extend this algorithm to produce truly Delaunay meshes for any PLC. First, as in my recently published 2D algorithm, I split the input line segments such that the length of the subsegments is asymptotically proportional to the local feature size (LFS). For each facet, I refine the mesh such that the edge lengths and the radius of the circumcircle of every triangular element are asymptotically proportional to the LFS. Finally, I refine the volume mesh to produce a constrained Delaunay mesh whose tetrahedral elements are well graded and have a radius-edge ratio less than some $\omega^* > 2/\sqrt{3}$ (except ``near'' small input angles). I extend this algorithm to generate truly Delaunay meshes by ensuring that every triangular element on a facet satisfies Gabriel's condition, i.e., its diametral sphere is empty. On an ``apex'' vertex where multiple facets intersect, Gabriel's condition is satisfied by a modified split-on-a-sphere (SOS) technique. On a line where multiple facets intersect, Gabriel's condition is satisfied by mirroring meshes near the line of intersection. The SOS technique ensures that the triangles on a facet near the apex vertex have angles that are proportional to the angular feature size (AFS), a term I define in the paper. All tetrahedra (except ``near'' small input angles) are well graded and have a radius-edge ratio less than $\omega^* > \sqrt{2}$ for a truly Delaunay mesh. The upper bounds for the radius-edge ratio are an improvement by a factor of $\sqrt{2}$ over current state-of-the-art algorithms. 
\end{abstract}

\pagebreak
\section{Introduction}

I present an algorithm to generate graded tetrahedral Delaunay meshes whose elements have an improved bound on their quality. I first present the algorithm to construct constrained Delaunay meshes and then extend it for truly Delaunay meshes. The algorithm is an extension of my recently published algorithm for 2D meshes~\cite{Sas21}, which may be viewed as a generalization of Chew's first algorithm~\cite{C89} for Delaunay mesh refinement. Dey et al.~\cite{DBS91} extended Chew's algorithm for 3D constrained Delaunay meshes. Accordingly, my 3D algorithm may also be viewed as a generalization of Dey et al.'s algorithm. 

The input to my algorithm is a piecewise linear complex (PLC), whose formal definition is provided in Section 2. A PLC is a generalization of a planar straight-line graph (PSLG) in 2D. For a truly Delaunay mesh in 2D, a triangular element does not contain any vertex inside its circumcircle. For a constrained Delaunay mesh, only vertices that are not ``visible'' are permitted inside a circumcircle. Visibility is occluded only by an input line segment.  In 3D, the same definition carries over, but the visibility may only be occluded by an input facet~\cite{C87, She02}. Ensuring that input facets are recovered by Delaunay tetrahedralization is a challenging aspect of generating truly Delaunay meshes. As a result, constrained Delaunay meshes are smaller and easier to generate than truly Delaunay meshes. A number of incremental techniques to compute the constrained or truly Delaunay tetrahedralization~\cite{JRS98, She08, SS14b, SS13} are known. 

Pioneered by Frey~\cite{Fre87}, Delaunay mesh refinement is typically carried out by adding Steiner vertices at a poor-quality triangle's circumcenter. The first provably good 2D constrained Delaunay mesh refinement algorithm was developed by Chew~\cite{C89}. In his algorithm, Chew first refines input line segments of a PSLG into subsegments such that their lengths are between some $h$ and $\sqrt{3}h$. Then, he constructs their constrained Delaunay triangulation. Next, triangles with radii of their circumcircle larger than $h$ are refined by adding their circumcenter.  These steps result in a uniform mesh (the areas of the triangles in the mesh are nearly the same) with the radius-edge ratio (the ratio of the radius of the circumcircle and the shortest edge) of triangles less than or equal to $1$. The radius-edge ratio is also the ``quality'' of an element. Chew~\cite{C93} extended his first algorithm by choosing to refine only those triangles whose radius-edge ratio is greater than a threshold. In this algorithm, if a potential vertex is too close to an existing subsegment, all Steiner vertices inside the diametral circle of the subsegment are deleted, and the midpoint of the subsegment is placed instead. Shewchuk~\cite{S97} showed that Chew's second algorithm produces a graded 2D mesh with a radius-edge ratio of triangles less than $\sqrt{5}/2 + \varepsilon$. In a graded mesh, the areas of triangles vary proportional to the distance to the two nearest input features. 

Dey et al.~\cite{DBS91} extended Chew's first algorithm for constrained Delaunay meshes in 3D by similarly refining the volume mesh. Their algorithm produces uniform meshes whose radius-edge ratio is less than or equal to $2$. Shewchuk~\cite{S97, S00, S02} extended Chew's second algorithm and developed a 3D constrained Delaunay mesh refinement algorithm, which produces graded meshes with tetrahedral elements whose radius-edge ratio is less than $2\sqrt{(2/3)} + \varepsilon$. Shewchuk proved the existence of a constrained Delaunay tetrahedralization for PLCs augmented with Steiner vertices if the diametral spheres of the subsegments are empty~\cite{She98}. He also presented an extension of the algorithm (by carefully subdividing input segments, when necessary) that improved the radius-edge ratio upper bounds to $2\sqrt{3}$, but the grading guarantee could not be extended in theory (but works in practice). Si~\cite{S09} improved the analysis of the algorithm and developed new algorithms to adaptively refine constrained Delaunay meshes. Si and Shewchuk~\cite{SS14} extended their algorithm for PLCs with small angles. Their algorithm ensures that the radius-edge ratio of elements away from a small angle is at most $2\sqrt{(2/3)} + \varepsilon$. 

Ruppert~\cite{R93, R95} developed an algorithm to generate a truly Delaunay, graded 2D mesh that continually inserts the circumcenters of poor-quality triangles. The algorithm does not allow the insertion of Steiner vertices inside the diametral circle of an input segment (or a subsegment). Instead, the midpoint of the segment (or the subsegment) is inserted. As diametral circles of subsegments are empty (called Gabriel's condition), these segments are recovered by Delaunay triangulation. Gabriel's condition is a sufficient condition to ``recover'' input segments by Delaunay triangulation, but it is not necessary. The recovery of input segments and facets makes truly Delaunay meshing tougher than constrained Delaunay meshing. If the input angles are greater $\pi/2$, with enough refinement and a conservative radius-edge bound requirement, it is possible to show that the constrained Delaunay algorithms above also produce a truly Delaunay mesh. Hudson et al.~\cite{Hud07,HMP06}, Li et al.~\cite{LTU99} and Shewchuk~\cite{S97} developed algorithms that can construct meshes whose upper bound on their elements' radius-edge ratio is at least $2 + \varepsilon$.

Algorithms that generate truly Delaunay mesh refinement in 3D typically ensure Gabriel's condition (empty diametral spheres) is satisfied even when multiple facets or line segments meet at a small angle. Murphy et al.~\cite{MMG00} developed the first algorithm that generates a truly Delaunay mesh for 3D domains with small angles, but no grading or quality guarantees were provided. In their algorithm, they ``protect'' regions near input vertices (where multiple facets and segments meet) and refine segments (where multiple facets meet) by carefully placing a layer of vertices around them. This strategy ensures that facets are recovered near those features. Away from those features, the facets and segments are fine enough that the diametral spheres do not intersect their adjacent facets or segments. Thus, Gabriel's condition is satisfied everywhere. After the facet meshing, the volume mesh refinement uses circumcenter insertion. Although their algorithm is not practical, this strategy of protecting input vertices and segments has been used in almost every algorithm since. 

Cohen-Steiner et al.~\cite{CVY02} introduced new techniques to protect regions around input vertices and segments. To protect regions around vertices, they consider a set of small concentric spheres around a vertex and the intersection of the spheres with facets adjacent to the vertex. In each of the resulting concentric sectors, they place vertices on the circumference until Gabriel's condition is satisfied everywhere. Cohen-Steiner et al. called this the split-on-a-sphere (SOS) strategy, and this strategy has been used in all algorithms since. To protect the regions around segments, they carefully refine the segments and do not allow Steiner vertices to be placed close to the vertices on the subsegments. This algorithm is practical to be implemented, but it does not provide any quality or grading guarantees. 

Cheng and Poon~\cite{CP03} developed an algorithm that provided quality and grading guarantees, but it was not practical due to its requirement of explicitly computing the intersection of spheres around every pair of vertices on a refined segment. These spheres serve as protection regions around input segments and vertices. Their algorithm guarantees a radius-edge ratio of all tetrahedra (except the ones near small angles) to be less than $16 + \varepsilon$. 

Cheng et al.~\cite{CDRR05} developed and implemented an algorithm that improved the radius-edge upper bound to $2 + \varepsilon$ with grading guarantees. Cheng et al. do not protect edges, and as a result, their algorithm works only for manifold surfaces (only two facets are adjacent to an input segment).  They do protect vertices using the SOS strategy. Since they deal with manifold surfaces, the satisfaction of Gabriel's condition is not critical to recovering the facets. With enough refinement, Delaunay tetrahedralization recovers all facets. 

Pav and Walkington~\cite{PW05} retained a radius-edge bound of $2 + \varepsilon$ with their algorithm, which also works for non-manifold PLCs. Pav and Walkington use ``caps'', ``collars'', and ``lenses'' to protect input segments. Rand and Walkington~\cite{RW09} generalized the techniques of Murphy et al.~\cite{MMG00} and Cohen-Steiner et al.~\cite{CVY02} of protecting vertices and segments using ``collars`` and developed an algorithm that provides a grading guarantee and a radius-edge ratio bound of $2 + \varepsilon$. They also generalized Cheng and Poon's~\cite{CP03} strategy using ``intestines'', and the generalized algorithm provides a grading guarantee and a radius-edge ratio bound of $4 + \varepsilon$. Notably, they implemented both these generalization strategies. 

The algorithms above place Steiner vertices at the circumcenter of a poor-quality tetrahedron. Instead, one can also place them at off-center points such that they are sufficiently far away from other vertices in the mesh. {\"U}ng{\"o}r et al.~\cite{U04, U09, EU09,RU08}, Chernikov and Chrisochoides~\cite{CC09, FCC10, CC12}, Rivara et al.~\cite{RRP18, RR14, RD20}, Hudson~\cite{Hud07,H08}, and Engwirda~\cite{E15, E15a} developed heuristic algorithms that place Steiner vertices at off-center points. They have shown that their algorithms provide quality guarantees that are as good as the algorithms above. In practice, some of these algorithms generate meshes that are smaller than the ones discussed before. 

In my previous paper~\cite{Sas21}, I developed an algorithm to generate both constrained and truly Delaunay 2D meshes. I then refine an input segment such that the length of the subsegments is proportional to the local feature size (LFS) at their endpoints. I refine the resulting triangulation by prioritizing poor-quality triangles with the shortest edges first and placing their circumcenter or an off-center point. This algorithm generates graded meshes with a radius-edge ratio upper bound less than $1 + \varepsilon$. 

In this paper, I use a similar strategy to generate constrained Delaunay meshes in 3D. The segment refinement algorithm is identical to the 2D algorithm. Next, the facet refinement is carried out such that the radius of the circumcircle and length of the edges of a triangle are proportional to the LFS. Next, the volume refinement places the circumcenter or an off-center point of a poor-quality tetrahedron. As before, I prioritize tetrahedra with the shortest edges first. I will show that my algorithm produces meshes whose elements have a radius-edge ratio less than $2/\sqrt{3} + \varepsilon$.

I extend the algorithm above to generate truly Delaunay meshes. I use a modified SOS strategy to protect input vertices and refine the sectors such that their angles are proportional to the angular feature size (AFS), which is defined in Section 5. Instead of protecting a segment of the intersection of multiple facets, I mirror the facet mesh close to the segment. This extension results in a graded mesh whose elements have a radius-edge ratio less than $\sqrt{2} + \varepsilon$.

\section{Background}

In this section, I define common terms and notations used in the paper. I also briefly describe 
my 2D advancing-front triangular mesh generation algorithm~\cite{Sas21} since the algorithms in this paper 
are an extension of the algorithm presented there. Although I provide relevant details
and the statement of relevant lemmas in this paper, I strongly recommend the reader be 
familiar with that paper. 

\subsection{Piecewise Linear Complex (PLC)}

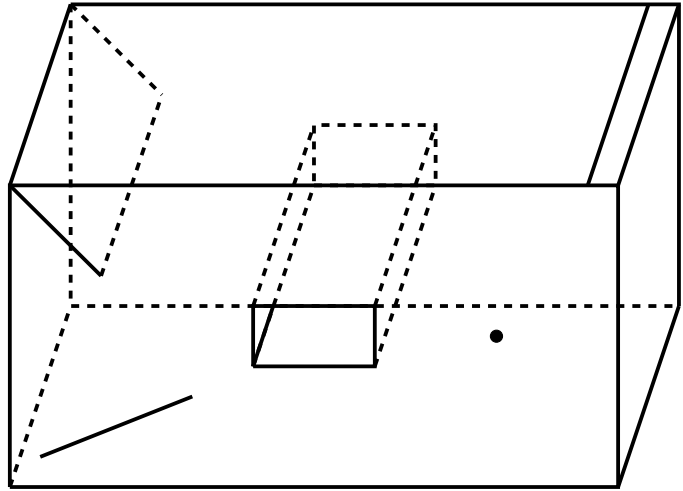
\begin{figure}
\centering
\begin{tikzpicture}[scale=4]
	\draw [line width=0.5mm] (0,0) -- (2,0) -- (2,1) -- (0,1) -- (0,0);
	\draw [line width=0.5mm, dashed] (0.2,1.6) -- (0.2,0.6) -- (2.2,0.6);
	\draw [line width=0.5mm] (2.2,0.6) -- (2.2,1.6) -- (0.2,1.6);
	\draw [line width=0.5mm] (0,1) -- (0.2,1.6);
	\draw [line width=0.5mm] (1.9,1.0) -- (2.1,1.6);
	\draw [line width=0.5mm] (2.0,0.0) -- (2.2,0.6);
	\draw [line width=0.5mm] (2.0,1.0) -- (2.2,1.6);
	\draw [line width=0.5mm, dashed] (0.0,0.0) -- (0.2,0.6);

	\draw [line width=0.5mm] (0,1) -- (0.3,0.7);
	\draw [line width=0.5mm, dashed] (0.2,1.6) -- (0.5,1.3);
	\draw [line width=0.5mm, dashed] (0.3,0.7) -- (0.5,1.3);

	\draw [line width=0.5mm] (0.8, 0.4) -- (1.2,0.4) -- (1.2, 0.6) -- (0.8, 0.6) -- (0.8,0.4);
	\draw [line width=0.5mm, dashed] (1.0, 1.0) -- (1.4,1.0) -- (1.4, 1.2) -- (1.0, 1.2) -- (1.0,1.0);
	\draw [line width=0.5mm, dashed] (0.8, 0.4) -- (1.0,1.0);
	\draw [line width=0.5mm] (0.8, 0.4) -- (0.866,0.6);
	\draw [line width=0.5mm, dashed] (1.2, 0.4) -- (1.4,1.0);
	\draw [line width=0.5mm, dashed] (1.2, 0.6) -- (1.4,1.2);
	\draw [line width=0.5mm, dashed] (0.8, 0.6) -- (1.0,1.2);

        \draw [fill=black] (1.6,0.5) circle [radius=0.02];
	\draw [line width=0.5mm] (0.1,0.1) -- (0.6,0.3);

\end{tikzpicture}
\caption{An example of a piecewise linear complex (PLC). Note the 
	stray line and a vertex on the facet closest to the reader and the
	nonmanifold facets.}
\label{fig:plc}
\end{figure}

A piecewise linear complex (PLC) is input for my meshing algorithm. As I focus on 3D algorithms, I will define the PLC for three dimensions. It is a set of vertices, edges, and facets that define a closed domain to be meshed. Edges are line segments joining pairs of input vertices and do not intersect anywhere else. Facets are planar polygons (with holes, possibly) bounded by edges, and facets intersect only along input edges or vertices. Edges may also lie on a facet or intersect with the facet at an input vertex. The domain may have 3D holes and stray vertices, edges, or facets. See Fig.~\ref{fig:plc} for an example. 

\subsection{Local Feature Size}

\begin{figure}
\centering
\begin{subfigure}[b]{0.3\textwidth}
\begin{tikzpicture}[scale=2]
    \draw [smooth,samples=100,domain=0:1] plot({\x},{sqrt((0.3-(\x))*(0.3-(\x)) + 0.2*0.2)});
    \draw [line width=0.75mm, blue] (0,0) -- (1,0);
    \draw [fill=red] (0.3,0.2) circle [radius=0.05cm];
    \node [below left] at (0,0) {p};
    \node [below right] at (1,0) {q};
\end{tikzpicture}
\caption{Vertex function}
\end{subfigure}
\centering
\begin{subfigure}[b]{0.3\textwidth}
\begin{tikzpicture}[scale=2]
    \draw [line width=0.75mm, blue] (0,0) -- (1,0);
    \draw [line width=0.25mm] (0.25,0.4) -- (0.75,0.2);
    \draw [line width=0.25mm, red] (0.45,0.3464) -- (0.85,0.1732);
    \draw [line width=0.25mm, dashed] (0.45,0.3464) -- (0.25,0.0);
    \draw [line width=0.25mm, dashed] (0.85,0.1732) -- (0.75,0.0);
    \draw [line width=0.25mm, dotted] (0.25,0.0) -- (0.25,0.4);
    \draw [line width=0.25mm, dotted] (0.75,0.0) -- (0.75,0.2);
    \node [below left] at (0,0) {p};
    \node [below right] at (1,0) {q};
    \node [above] at (0.45,0.3464) {c};
    \node [above] at (0.85,0.1732) {d};
    \node [below] at (0.25,0) {$x_c$};
    \node [below] at (0.75,0) {$x_d$};
\end{tikzpicture}
\caption{Line function}
\end{subfigure}
\centering
\begin{subfigure}[b]{0.3\textwidth}
\begin{tikzpicture}[scale=2]
    \draw [line width=0.75mm, blue] (0,0) -- (1,0);
    \draw [line width=0.1mm,  black] (0,1) -- (0.5,0.5); 
    \draw [line width=0.1mm,  black] (0.5,0.5) -- (1,1);
    \draw [fill=red] (0.0,0.0) circle [radius=0.05cm];
    \draw [fill=red] (1.0,0.0) circle [radius=0.05cm];
    \node [below left] at (0,0) {p};
    \node [below right] at (1,0) {q};
\end{tikzpicture}
\caption{Augmented function}
\end{subfigure}
\caption{The various distance functions associated with 
a line segment $pq$ in the PLC. The blue, thick line segment $pq$
is horizontal and can be considered as part of the $x$ axis with
vertex $p$ being at the origin.  The distance to the red feature(s) 
is a function of $x$, and the function is plotted as a thin black curve. 
(a) The distance to a PLC vertex is plotted as a function of $x$.
The domain of the function is from $p$ to $q$.
(b) The distance to some other PLC line segment (red) or a plane 
is plotted. The distance varies linearly, and  
the domain of the linear distance function is limited.  
The dashed lines are perpendicular to the PLC line segment (red) or plane, 
and they limit the domain of the distance function.  
Beyond the domain, the distance to $c$ or $d$ (whichever is closer) 
defines the distance function on $pq$.  Those parts of the distance 
functions look like the distance function in (a). 
(c) As $p$ and $q$ are not adjacent,
the distance to $p$ or $q$, whichever is larger, also limits the 
feature size at any point on the line segment $pq$. A disk centered 
at a point on $pq$ with the radius equal to the greater of 
$xp$ or $xq$ contains both $p$ and $q$. Thus, this piecewise
linear function is also considered to compute the local
feature size.}
\label{fig:distance}
\end{figure}
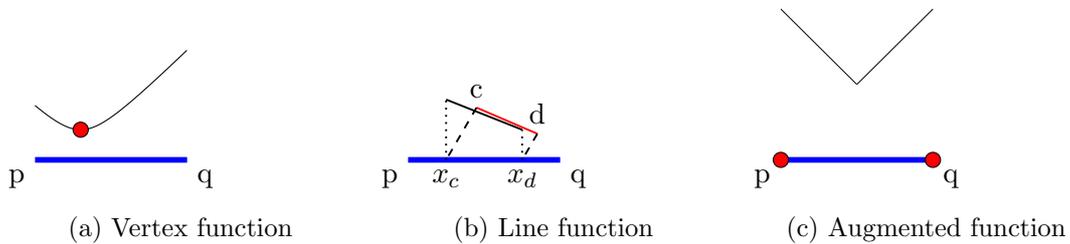

The local feature size (LFS) is defined as the largest sphere that can be drawn at a point such that only one set of adjacent features is present on or in the sphere. A feature is a vertex, line segment, or facet. For a given point, the LFS is equal to the distance to the second nearest nonadjacent feature from the point. Note that an input line segment and any one of its endpoints are adjacent features, whereas the two endpoints are nonadjacent features. For a given point on an input line segment, the distance to the nearest nonadjacent feature is the LFS at that point. Fig.~\ref{fig:distance} provides a possible set of such distance functions. The lower envelope of these distance functions for a PLC line segment defines the LFS function for the line segment. 

In 2D, if the lengths of all mesh edges are less than $\gamma \mathrm{LFS}(x)$, where $\gamma$ is a constant and $x$ is one of the endpoints of the edge, the mesh is only a constant times larger than the smallest mesh possible with the same radius-edge ratio requirements~\cite{R95}. This result does not extend to 3D, but it implies that the mesh is well-graded away from small features. 

If small angles (defined below) are present in the input PLC, the LFS may not be a useful metric ``near'' the small angles. If the PLC has small angles, Shewchuk~\cite{S97,S00} showed that it is impossible to construct a mesh such that the small angle is confined to the input features alone, i.e, other small angles will be present ``near'' the input feature. In such cases, the bound on the length of the edges cannot be satisfied. In my algorithms (in the previous paper~\cite{Sas21} and this paper), the lengths of the edges near the small-angle are dictated by how small the angle angle is.

\begin{figure}
\centering
\begin{tikzpicture}[scale=2]
	\draw [line width=0.30mm] (0,0) -- (1,0) -- (1,1) -- (-1,1) -- (-1, -0.75) -- (-0.5, -0.75) -- (-0.5, 0.5) -- (0, 0.5) -- (0,0);
	\draw [line width=0.30mm] (0,0) -- (1,0) -- (1,-1) -- (-1.5,-1) -- (-1.5, -0.25) -- (-1, -0.25);
	\draw [line width=0.30mm] (-0.5, -0.25) -- (0, -0.25) -- (0, 0);
	\draw [line width=0.30mm, dotted] (-1, -0.25) -- (-0.5, -0.25);
\end{tikzpicture}
\caption{
Even when the angle between two facets at the line of intersection is
close to 180 degrees, parts of the two
facets may be at a much smaller angle due to the nonconvex geometry.
}
\label{fig:nonconvexsmallangle}
\end{figure}
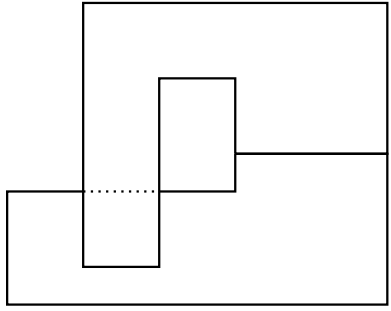

In 3D, two adjacent facets may form a large dihedral angle of $\pi - \epsilon$ (see Fig.~\ref{fig:nonconvexsmallangle}), but due to their nonconvex geometry, parts of the facets may be at an angle of $\epsilon$ from each other. In such cases, although the LFS at some points on the facets is large, the length of the mesh edges and the quality of elements near the feature are dictated by the small angle. 

\subsection{Asymptotically Proportional Edge Length}
In my previous paper, I split the input line segments into multiple subsegments such that their lengths are bounded from above and below as a proportion of the LFS. The constants of proportionality for the lower and upper bounds are, of course, different. Mathematically, 
$$ A^* \le \frac{\mathrm{LFS(x)}}{l_x} \le B^*,$$
where $A^*$ and $B^*$ are some constants, $\mathrm{LFS(x)}$ is the LFS at some vertex $x$, and $l_x$ is the length of a subsegment adjacent to $x$. In my 2D algorithm~\cite{Sas21}, as I refine the segments more and more, the value of $A^*$ and $B^*$ increase, but their ratio, $B^*/A^*$, tends to 1. I use the algorithm in this paper, too. In addition to subsegments on an input line segment, I ensure that the lengths of edges on an input facet are also asymptotically proportional to the LFS at their endpoints. The constants associated with the proportionality are $A^{**}$ and $B^{**}$. Clearly, $A^{**} \le A^* \le B^* \le B^{**}$. 

\subsection{Small Angle}

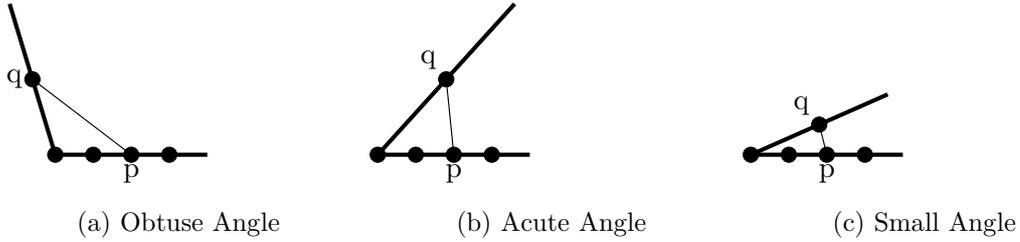
\begin{figure}
\centering
\begin{subfigure}[b]{0.3\textwidth}
\begin{tikzpicture}[scale=2]
    \draw [line width=0.55mm] (0,0) -- (1,0);
    \draw [line width=0.55mm] (0,0) -- (-0.3,1);
    \draw [line width=0.15mm] (0.5,0) -- (-0.15,0.5);
    \draw [fill=black] (0,0) circle (0.05);
    \draw [fill=black] (0.5,0) circle (0.05);
    \draw [fill=black] (0.25,0) circle (0.05);
    \draw [fill=black] (0.75,0) circle (0.05);
    \draw [fill=black] (-0.15,0.5) circle (0.05);
    \node [below] at (0.5,0) {p};
    \node [left] at (-0.15,0.5) {q};
\end{tikzpicture}
\caption{Obtuse Angle}
\end{subfigure}
\centering
\begin{subfigure}[b]{0.3\textwidth}
\begin{tikzpicture}[scale=2]
    \draw [line width=0.55mm] (0,0) -- (1,0);
    \draw [line width=0.55mm] (0,0) -- (0.9,1);
    \draw [line width=0.15mm] (0.5,0) -- (0.45,0.5);
    \draw [fill=black] (0,0) circle (0.05);
    \draw [fill=black] (0.5,0) circle (0.05);
    \draw [fill=black] (0.25,0) circle (0.05);
    \draw [fill=black] (0.75,0) circle (0.05);
    \draw [fill=black] (0.45,0.5) circle (0.05);
    \node [below] at (0.5,0) {p};
    \node [above left] at (0.45,0.5) {q};
\end{tikzpicture}
\caption{Acute Angle}
\end{subfigure}
\centering
\begin{subfigure}[b]{0.3\textwidth}
\begin{tikzpicture}[scale=2]
    \draw [line width=0.55mm] (0,0) -- (1,0);
    \draw [line width=0.55mm] (0,0) -- (0.9,0.4);
    \draw [line width=0.15mm] (0.5,0) -- (0.45,0.2);
    \draw [fill=black] (0,0) circle (0.05);
    \draw [fill=black] (0.5,0) circle (0.05);
    \draw [fill=black] (0.25,0) circle (0.05);
    \draw [fill=black] (0.75,0) circle (0.05);
    \draw [fill=black] (0.45,0.2) circle (0.05);
    \node [below] at (0.5,0) {p};
    \node [above left] at (0.45,0.2) {q};
\end{tikzpicture}
\caption{Small Angle}
\end{subfigure}
\centering
\caption{
Images depicting the problem with small angles
in a planar straight-line graph (PSLG), which is what a PLC 
in 2D is called. 
The thick lines are part of the PSLG. 
The black dots are the vertices added to the 
PSLG (not all are shown).
(a) When the angle is obtuse, the line segment
$pq$ is longer than subsegments at $p$ and $q$. (b) When the
angle is acute, but greater than 60\degree, $pq$ might be
longer than subsegments at $p$ and $q$, but it is not
guaranteed unless the segments are adequately refined. 
(c) When the angle
is very small, there is a good chance that $pq$ is
shorter than the threshold for size optimality.  
}
\label{fig:smallangleproblem}
\end{figure}

In my previous paper, I defined a small angle between two adjacent line segments as an angle $\phi < \arccos{(\frac{1}{2R})}$, where $R = B^*/A^*$ ($A^*$ and $B^*$ are defined above). I use the same definition in this paper, too. Also, for dihedral angles between two adjacent facets or a line segment and a facet, any angle less than $\pi/2$ is considered a small angle. The reason these angles are considered small is that the length of an edge joining vertices in the adjacent facets or edges may be smaller than what is desired by the asymptotic proportionality requirements. This reasoning is explained in Fig.~\ref{fig:smallangleproblem} for 2D meshes. 

\subsection{Skinny Triangle}

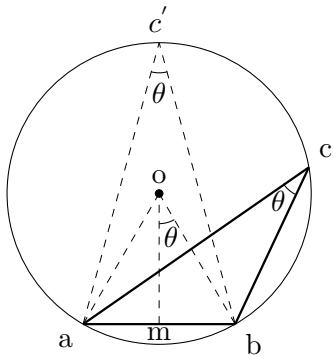
\begin{figure}
\centering
\begin{tikzpicture}[scale=2]
    \draw (0,0) circle (1);
    \draw [line width=0.30mm] (240:1) -- (300:1);
    \draw [line width=0.30mm] (300:1) -- (10:1);
    \draw [line width=0.30mm] (240:1) -- (10:1);
    \draw [line width=0.10mm, dashed] (300:1) -- (90:1);
    \draw [line width=0.10mm, dashed] (240:1) -- (90:1);
    \draw [fill=black] (0,0) circle (0.025);
    \draw [line width=0.10mm, dashed] (300:1) -- (0:0);
    \draw [line width=0.10mm, dashed] (240:1) -- (0:0);
    \draw [line width=0.10mm, dashed] (270:0.866025) -- (0:0);

    \draw  (90:1) ++(255:0.2) arc (255:285:0.2);
    \draw  (10:1) ++(213:0.2) arc (213:243:0.2);
    \draw  (0:0) ++(270:0.2) arc (270:300:0.2);

    \node [above] at (0,0) {o};
    \node [below left] at (240:1) {a};
    \node [below right] at (300:1) {b};
    \node [below] at (270:0.82){m};
    \node [above right] at (10:1) {c};
    \node [above] at (90:1) {$c^{'}$};
    \node [below] at (90:0.8) {$\theta$};
    \node at ($(10:1)+(228:0.3)$) {$\theta$};
    \node at (285:0.3) {$\theta$};
\end{tikzpicture}
\caption{
The relationship between the minimum angle in a triangle
and the ratio of the radius of its circumcircle and the
length of the shortest edge. Let $ab$ be the shortest
edge of $\triangle abc$.  Then, the angle at $c$ is its 
shortest angle.  The triangle's circumcircle is shown. 
Note that $\angle c^{'} = \angle c$. Also, $\angle aob = 
2\angle c^{'}$, and $\angle mob = \angle c^{'}$.  Clearly,
$\sin{\angle c} = \sin{\angle c^{'}} = 
\sin{\angle mob} = \frac{|ab|}{2|ob|}
= \frac{l}{2r}$, where $l = |ab|$ and $r$ is the length
of the radius of the circumcircle. 
}
\label{fig:angleratio}
\end{figure}

In the Delaunay refinement literature, a common metric used to determine the quality of a triangle (or a simplex any dimension) is the radius-edge ratio, which measures the ratio of the radius of the circumscribing circle and the shortest edge in the triangle. If this ratio is larger than some $\alpha^*$, it is a skinny triangle. Fig.~\ref{fig:angleratio} explains how the radius-edge ratio is related to the minimum angle in the triangle. 

If a triangle's shortest edge joins two vertices on two adjacent line segments separated by a small angle, I slightly modify the definition of a skinny triangle. I consider the apparent length of its shortest edge to be $l_a = \mathrm{max}(l, f/B^*)$, where $l$ is the length of the real shortest edge and $f$ is the LFS at one of its endpoints. If the radius circumsphere of the triangle is $r > l_a\alpha^*$, I consider it a skinny triangle. The reason for considering the apparent length instead of the actual length of the shortest side is that the shortest side may be too short when the angle is small, as illustrated in Fig.~\ref{fig:smallangleproblem}

\subsection{Skinny Tetrahedron}

If a tetrahedron's radius-edge ratio is less than a certain threshold $\omega^*$, it is a skinny tetrahedron. Although the dihedral angles or solid angles in a tetrahedron and the radius-edge ratio have no relationship, the metric is still useful in differentiating most good-quality tetrahedra from others. Unfortunately, this metric does not identify a sliver as a poor-quality tetrahedron. Nevertheless, it is still useful in analyzing Delaunay refinement algorithms. Also, postprocessing steps typically get rid of most slivers. As in the case of skinny triangles, I will make a similar exception for a skinny tetrahedron that is adjacent to a small angle. 

\subsection{Boundary Encroachment}

In many prior algorithms, the concept of vertex encroachment upon a boundary line segment or a facet was used to prevent poor-quality mesh elements near the boundary. In those algorithms, if any potential vertex came too close to a boundary features such that it is inside a feature's diametral sphere (for truly Delaunay meshes) or close enough to create poor-quality elements (for constrained Delaunay meshes), the potential vertex is not inserted into the mesh. In algorithms in my prior paper and this paper, I show that no vertex encroaches upon a boundary feature if $\alpha^*$ and $\omega^*$ are chosen carefully. 

\subsection{The LFS-Edge Ratio}
\label{sec:lfsedge}
The LFS-edge ratio is the ratio of the LFS at a vertex and the length of a mesh edge adjacent to the vertex. For a mesh of a facet, the LFS-edge ratio is bounded from below by some $A^{**}$ and from above by some $B^{**}$. If the desired radius-edge ratio is $\alpha^* > 1$, it is easy\footnote{
When the shortest edge is of length $1$ unit, the largest radius of the circumcircle of a nonskinny triangle is $\alpha^*$ units. The longest edge may be as long as the diameter, which is $2\alpha^*$ units. 
} to show that $B^{**} = 2\alpha^* A^{**}$. 

\subsection{2D Advancing-Front Delaunay Mesh Refinement Algorithm}

Here, I will briefly describe the algorithm in the previous paper. The input to the algorithm is a 2D PLC, which is also called a planar straight-line graph (PSLG), and the desired radius-edge ratio $\alpha^*$. The output is a mesh whose elements have the desired radius-edge ratio except ``across'' the small angles in the input. The algorithm consists of three steps:
\begin{enumerate}
\item Compute the LFS along every input line segment. 
\item Refine every line segment such that the subsegments are proportional to the LFS.
\item Refine the triangulation using ``off-centers'' or circumcenters of skinny triangles. 
\end{enumerate}

I will describe each of the steps in more detail below.

\subsubsection{Computing the LFS}

The LFS at a point on an input line segment is the radius of the largest circle that does not intersect a nonadjacent feature or the radius of the largest circle that does not enclose both its endpoints, whichever is smaller. Thus, the LFS along an input line segment is the lower envelope of distance functions from the line segment to nonadjacent features and the distance function to the farthest endpoint. Quadtrees, Voronoi diagrams of input line segments, and kinetic hangers may be used to accelerate the computation of the LFS function along an input segment. 

\subsubsection{1D LFS-Based Refinement}

\begin{figure}
  \centering
  \begin{tikzpicture}
        \draw[thick] (0,0) -- (10,0);
        \draw[thick] (1,2) -- (11,2);
        \draw[-> {triangle 45},semithick,red] (0,0) -- (1,2);
        \draw[-> {triangle 45},semithick,red] (2,0) -- (5,2);
        \draw[-> {triangle 45},semithick,red] (4,0) -- (8,2);
        \draw[-> {triangle 45},semithick,red] (6,0) -- (9.5,2);
        \draw[-> {triangle 45},semithick,red] (8,0) -- (10.5,2);
        \draw[-> {triangle 45},semithick,red] (10,0) -- (11,2);
        \draw[semithick] (0,-0.1) -- (0,0.1);
        \draw[semithick] (2,-0.1) -- (2,0.1);
        \draw[semithick] (4,-0.1) -- (4,0.1);
        \draw[semithick] (6,-0.1) -- (6,0.1);
        \draw[semithick] (8,-0.1) -- (8,0.1);
        \draw[semithick] (10,-0.1) -- (10,0.1);
        \draw[semithick] (1,1.9) -- (1,2.1);
        \draw[semithick] (5,1.9) -- (5,2.1);
        \draw[semithick] (8,1.9) -- (8,2.1);
        \draw[semithick] (9.5,1.9) -- (9.5,2.1);
        \draw[semithick] (10.5,1.9) -- (10.5,2.1);
        \draw[semithick] (11,1.9) -- (11,2.1);
  \end{tikzpicture}
  \caption{An example of reference-to-PSLG mappings $M_i(t)$ from a
    reference segment $T_i$ to a PSLG segment $L_i$. 
    The reference segment is uniformly split into $n$ subsegments, and the
    corresponding splits are made in the PSLG segment.  The mapping function
    is defined such that uniform splits in the reference segment
    correspond to asymptotically proportional (to the local feature size) 
    splits in the PSLG segment.  Note that the reference segment and
    the PSLG segment may be of different lengths.}
  \label{fig:map}
\end{figure}
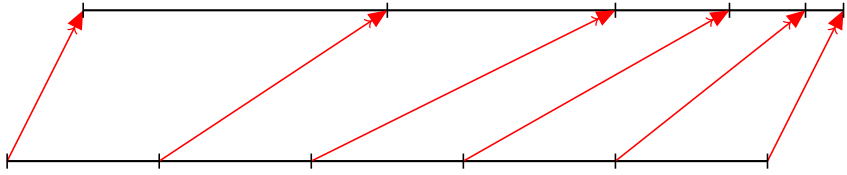

To split a line segment, I first construct a reference segment, and I map every point on this reference segment to a point on the line segment (see Fig.~\ref{fig:map}). I obtain the mapping function by solving the following differential equation $M^{'}(t) = F(M(t))$, where $M$ is the mapping function and $F$ is the local feature size function. Solving the differential equation with the appropriate boundary conditions also provides the length of a reference segment. I consider the shortest reference segment among the reference segments for all the line segments in the input geometry and divide it into $n^*$ parts, where $n^*$ is determined as a function of the desired radius-edge ratio (smaller the desired radius-edge ratio, greater the value of $n^*$) and the smallest angle (smaller the angle, greater the value of $n^*$). Other segments are divided into $n = n^* \floor{T/T_{\mathrm{min}}}$, where T is the length of its reference segment, and $T_{\mathrm{min}}$ is the length of the shortest reference segment above. More details about this algorithm are provided in my previous paper~\cite{Sas21}. 

\subsubsection{The 2D Advancing-Front Delaunay Refinement}

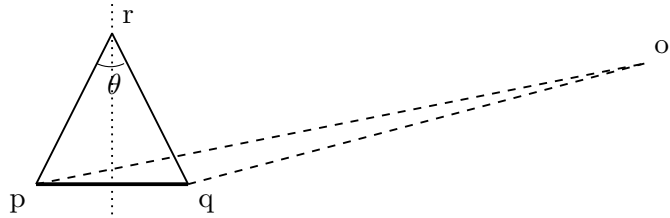
\begin{figure}
\centering
\begin{tikzpicture}[scale=2]
    \draw [line width=0.50mm] (0,0) -- (1,0);
    \draw [line width=0.25mm] (0,0) -- (0.5,1);
    \draw [line width=0.25mm] (1,0) -- (0.5,1);
    \draw [line width=0.25mm, dashed] (0,0) -- (4,0.8);
    \draw [line width=0.25mm, dashed] (1,0) -- (4,0.8);
    \draw [line width=0.25mm, dotted] (0.5,-0.2) -- (0.5,1.2);
    \draw (0.4,0.8) arc (243.44:296.56:0.2);
    \node [below left] at (0,0) {p};
    \node [below right] at (1,0) {q};
    \node [above right] at (0.5,1) {r};
    \node [above right] at (4,0.8) {o};
    \node [below right] at (0.4,0.8) {$\theta$};
\end{tikzpicture}
\caption{The off-center vertex. Instead of the circumcenter of 
a skinny triangle $pqo$, \"{U}ng\"{o}r and Erten~\cite{EU09} add an 
off-center vertex into the Delaunay triangulation.  The off-center
vertex $r$ lies on the perpendicular bisector of the shortest edge
$pq$ of the skinny triangle such that the edge subtends the minimum
desired angle at the point.  Note that $r$ should be on the same
side of $pq$ as $o$ is. As in their algorithm, I use the same
point for Delaunay refinement if it is closer to the shortest
edge than the circumcenter.}
\label{fig:offcenter}
\end{figure}
After the input line segments are refined, I prioritize skinny triangles with the shortest edge first and refine the triangulation by placing the skinny triangle's off-center vertex (see Fig.~\ref{fig:offcenter}) or the circumcenter (whichever is closer). I ignore skinny triangles whose vertices on the shortest edge lie on adjacent input line segments separated by a small angle (see the modified definition of a skinny triangle within small angles in Section 2.5 above). As in all other algorithms, I recompute the constrained or truly Delaunay triangulation after every vertex is added. I have proven that this algorithm terminates with a size-optimal mesh if $\alpha > 1$ (except ``across'' a small angle). Since the Steiner vertices are never placed too far from existing vertices in the mesh, the algorithm behaves like an advancing-front mesh refinement algorithm.

\section{The 3D Constrained Delaunay Refinement Algorithm}
I will now describe how my 2D advancing-front algorithm may be extended to 3D for 
constrained Delaunay meshes. The input to the algorithm is a PLC and the 
desired radius-edge ratio $\omega^*$.
The algorithm consists of four steps:
\begin{enumerate}
\item Compute the LFS along every input line segment. 
\item Refine every line segment such that the subsegments are proportional to the LFS.
\item Refine the facet triangulation using ``off-centers'' or circumcenters of skinny triangles until 
	the radii of the circumcircles of the triangles are asymptotically proportional to the LFS.
\item Refine the tetrahedral mesh using ``off-centers'' or circumcenters of skinny tetrahedra by prioritizing the ones with the shortest edges first. 
\end{enumerate}
The first step has been described in detail in the previous work. I will describe the next three steps in detail below. Note that facet
refinement involves recomputation of constrained Delaunay triangulation on the facet using an incremental algorithm after each
vertex is added, and volume mesh refinement involves the recomputation of constrained Delaunay tetrahedralization.

\subsection{1D LFS-Based Refinement}

The algorithm to split an input segment into subsegments that are asymptotically proportional to the LFS at their endpoints is identical to the one described in my prior paper~\cite{Sas21}. That paper describes the rationale behind each of the steps in more detail. I have briefly described the algorithm in Section 2.6. The only unanswered question is how much the input line segments should be refined. The answer depends on the desired radius-edge ratio, $\omega^*$, of the tetrahedra in the final mesh. We will see in Section 4 that the triangular surfaces mesh elements on 2D facets of the PLC should also have a bounded (from above) radius-edge ratio, $\alpha^* = \frac{\sqrt{3}}{2}\omega^*$. Thus, we should pick $n^*$ (defined in Section 2.4) such that it guarantees the desired radius-edge ratio of the mesh elements on the facets. The detailed derivation of the equation that dictates the value of $n^*$ is in Section 4. Since a constrained Delaunay tetrahedralization is guaranteed to exist if the subsegments are strongly Delaunay~\cite{She98}, $n^*$ should also be large enough to satisfy that constraint. I have shown in my previous paper~\cite{Sas21} that there exists a minimum such $n^*$ even for domains with small angles. 

\subsection{2D LFS-Based Advancing-Front Refinement}

After the input line segments are split using the algorithm above, I compute the truly Delaunay triangulation on each facet and refine the triangulations independently. The goal of this refinement is to ensure that the lengths of all edges in the surface mesh are asymptotically proportional to the LFS at their endpoints. Also, the radius-edge ratio of all triangles should be less than $\alpha^*$, and consequently, the radius of the circumcircle should be asymptotically proportional to the LFS at the vertices. These objectives can be achieved by refining those triangles whose radius of the circumcircle is greater than a threshold (to be defined later in Lemma~\ref{lemma:bigalpha}) or whose radius-edge ratio is greater $\alpha^*$. As in prior work~\cite{Sas21}, I prioritize such triangles based on the length of their shortest edges and refine them by adding an off-center vertex (defined below) or the circumcenter (whichever is closer). When this algorithm terminates, we will have an LFS-constrained truly Delaunay triangulation on every facet. 

To define an off-center vertex, consider a triangle's shortest edge $l$ that joins vertices $a$ and $b$. On the perpendicular bisector of $l$, find any point $o$ such that $oa$ and $ob$ satisfy the LFS-based constraints on the edge size. This definition is different from the one shown in Fig.~\ref{fig:offcenter} because the angle subtended by $ab$ at $o$ is not considered. Note that off-center vertex insertion in the facet mesh is not necessary for constrained Delaunay refinement, but it is necessary for truly Delaunay triangulation. Heuristically, if the off-center vertex is as far away from other vertices as possible, the algorithm is likely to generate a smaller mesh. 

I will show in Section 4 that this algorithm never adds vertices that are too close to existing vertices if $\alpha^*$, $A^*$, $B^*$, $A^{**}$, and $B^{**}$ are chosen properly, and therefore, the algorithm terminates. I will also show that the algorithm never adds vertices that are too far from existing vertices either. Therefore, it retains the advancing-front behavior seen in prior work~\cite{Sas21}. This behavior ensures that triangles are as large as they can be while ensuring the LFS-based constraints. 

\subsection{3D Advancing-Front Refinement}

The 3D Delaunay refinement algorithm is similar to the 2D refinement algorithm in my previous paper~\cite{Sas21}. I prioritize skinny tetrahedra (whose radius-edge ratio is less than $\omega^*$) based on the length of their shortest edge (shorter ones first; longer ones later). Let the length of the shortest edge be $l$. I ignore most tetrahedra whose vertices on the shortest edge lie on adjacent input facets with a small dihedral angle (less than $\pi/2$), i.e., I use the modified definition of a skinny tetrahedra for such elements. For any other skinny tetrahedron, I consider its circumsphere with center $o$ and radius $r$. A Steiner vertex may be placed anywhere inside a concentric sphere (whose center is $o$) and radius $r - \omega^*l$, where $l$ is the length of its shortest edge. Any such point is at least a distance $\omega^*l$ from all other vertices that are not occluded by an input facet. I choose a point that is at most a distance of $\Gamma\omega^*l$, where $\Gamma \ge 1$ is some small constant ($\Gamma$ = 2, say), from one of the endpoints of the shortest edge of the tetrahedra (but still inside the concentric sphere). This choice ensures that new vertices are not too far away from existing vertices. This restriction also ensures that the algorithm behaves like an advancing front technique. More importantly, I will prove that this restriction also results in better bounds on edge lengths. Ideally, I will choose a point such that the resulting tetrahedra will have a radius-edge ratio of exactly $\omega^*$ (if it exists). If the radius-edge ratio of a resulting tetrahedra is greater than $\omega^*$, the resulting tetrahedra would get refined again. A ratio less than $\omega^*$ would create an unnecessarily small tetrahedron. Note that such a point will not exist if all of the tetrahedron's triangular facets have a radius-edge ratio greater than $\omega^*$. I still pick a point even if it results in a tetrahedron with a radius-edge ratio different from $\omega^*$. If the circumcenter of the tetrahedron is closer to one of the end points of the shortest edge of the tetrahedron than the ideal point, I add the circumcenter instead. 

\section{An Analysis of the Algorithm}

The analysis of the algorithm follows a similar template as in the prior work~\cite{Sas21}. I will show that the lengths of the edges in the mesh are bounded (from above and below) as a function of the LFS. I will show that such refinement is possible and no vertex encroaches upon a boundary feature if $A^*$, $B^*$, $A^{**}$, and $B^{**}$ are large enough for a chosen $\alpha^* > 1$ and $\omega^* > \frac{2}{\sqrt{3}}\alpha^*$.

\subsection{1D LFS-Based Refinement}
For input segment refinement, I have provided detailed proofs in the previous paper~\cite{Sas21} for the lemmas below. I will restate the relevant lemmas. These lemmas show that it is possible to split the input segments such that their lengths are asymptotically proportional to the LFS at their endpoints. 
They show that as the PLC segments are
progressively refined, the bounds $A^{*}$ and $B^{*}$ increase, but their
ratio $R$ approaches $1$. In addition, they show that given an upper
bound on $R$ or a lower bound on $A^{*}$, it is possible to split the PSLG
such that $B^{*}$ is bounded from above.

\begin{lemma}
\label{lemma:minsplit}
If the PSLG segment $L_i$ with the shortest reference segment
whose length is
$t^*_{\mathrm{min}}$
is split into $n^*$ subsegments, the bound on the ratio
of the local feature size and length of a subsegment
on $L_i$
at some vertex $p$ is given by
$A^*\le\frac{\mathrm{LFS}(p)}{l_p}\le B^*$, where
$l_p$ is the length of a subsegment one of whose endpoints
is $p$, $\mathrm{LFS}(\cdot)$ is the local feature size
function, and
$$A^* = \frac{n^*}{t^*_{\mathrm{min}}} - 1
\mathrm{\ and\ } 
B^* = \frac{n^*}{t^*_{\mathrm{min}}} + 1.$$
\end{lemma}

\begin{lemma}
\label{lemma:allsplit}
If the $i^\mathrm{th}$ PSLG segment with a
reference segment of length $t^*_i$
is split into $n = \floor{n^* \frac{t^*_i}{t^*_{\mathrm{min}}}}$
subsegments, the bound on the ratio
of the local feature size and length of the subsegment
at some vertex $p$ is given by
$A^*\le\frac{\mathrm{LFS}(p)}{l_p}\le B^*$, where
$l_p$ is the length of a subsegment one of whose endpoints
is $p$, $\mathrm{LFS}(\cdot)$ is the local feature size
function, and
$$A^* = \frac{n^*}{t^*_{\mathrm{min}}} - \frac{1}{2\log_e{2}}-1 
\mathrm{\ and\ } 
B^* = \frac{n^*}{t^*_{\mathrm{min}}} + 1.$$
\end{lemma}

\begin{lemma}
\label{lemma:aboundsb}
Given a lower bound $A^*$ on $\frac{\mathrm{LFS}(p)}{l_p}$,
it is possible to split the PSLG line segments such that the
upper bound $B^* \le A^* + 1/\log_e{2} + 2$, where $p$ is
a vertex on the PSLG line segment, and $l_p$ is
the length of a subsegment at $p$.
\end{lemma}

\begin{lemma}
\label{lemma:rboundsb}
Given an upper bound on the ratio $R>1$ of the upper and lower
bound on $\frac{\mathrm{LFS}(p)}{l_p}$, where $p$
is a vertex on the PSLG line segment and $l_p$ is
the length of a subsegment at $p$,
it is possible to split the PSLG line segments such that
an upper bound $B^*$ exists, and the bound 
is only a function of $R$.
\end{lemma}

\subsection{2D LFS-Based Advancing-Front Refinement}

In this section, I will show that it is always possible to refine a PSLG such that the lengths of the edges are asymptotically proportional to the LFS at their endpoints. I will also show that the radius of the circumcircle of a triangle is proportional to the LFS at its vertices.

I will first prove that the LFS-edge ratio differs by at most 1 for any edge when the LFS is measured at its two endpoints. For any edge that satisfies the LFS-edge ratio constraints, I will show that there  exists a point on its perpendicular bisector such that its distance to the edge's vertices satisfies the same constraints, which is the off-center vertex that may be inserted on a facet.

\begin{lemma}
\label{lemma:diffone}
Let $l_{ab}$ be the length of an edge with $a$ and $b$ as endpoints. If $\frac{\mathrm{LFS}(a)}{l_{ab}} = P$, $P - 1 \le \frac{\mathrm{LFS}(b)}{l_{ab}} \le P + 1$.
\end{lemma}
\begin{proof}
The LFS is a 1-Lipschitz function. This, $\mathrm{LFS}(b) \le \mathrm{LFS}(a) + l_{ab}$ and $\mathrm{LFS}(b) \ge \mathrm{LFS}(a) - l_{ab}$. The result follows.
\end{proof}

\begin{lemma}
If $A^{**} > 4$ and an edge $ab$ satisfies the LFS-edge constraints, there exists a vertex $v$ on the perpendicular bisector of $ab$ such that $va$ and $vb$ also satisfy the LFS-edge constraints.
\end{lemma}
\begin{proof}
Let $l$ be the length of $ab$ and $v$ be a point on the perpendicular bisector of $ab$. Let the lengths of $va$ and $vb$ be $\lambda l$, where $\lambda > 0$ is some constant. Let the LFS-edge ratio for edge $ab$ at $a$ be $P$, where $A^{**} \le P \le B^{**}$. The LFS-edge ratio of edge $va$ at $v$ is $\mathrm{LFS}(v) / |va|$, where $\mathrm{LFS}(v)$ is the LFS at $v$ and $|va| = \lambda l$. We have to find a $v$ such that $A^{**} \le \mathrm{LFS}(v) / |va| \le B^{**}$. Note that LFS-edge ratio has to be satisfied at $a$ and $b$ too. Thus, from the lemma above, we have to find a $v$ such that
$$A^{**} + 1 \le \frac{\mathrm{LFS}(v)} {|va|} \le B^{**} - 1.$$
Note that $\mathrm{LFS}(v) = \mathrm{LFS}(a) + \nu \lambda l$, where $-1 \le \nu \le 1$ (the LFS is a 1-Lipschitz function). Also note that $\mathrm{LFS}(a) = Pl$ and $|va| = \lambda l$. Thus, we have to find a $v$ such that 
$$A^{**} + 1 \le \frac{Pl + \nu \lambda l} {\lambda l} \le B^{**} - 1. $$ This inequality simplifies to finding $v$ such that 
$$A^{**} + 1 \le \frac{P}{\lambda} + \nu \le B^{**} - 1. $$ Since $-1 \le \nu \le 1$, the inequality further simplifies to 
$$A^{**} + 2 \le \frac{P}{\lambda} \le B^{**} - 2. $$
Note that $\lambda = 1/(2\cos{\theta})$, where $\theta = \angle bav$. Using this relation, we have to find a $v$ such that
	$A^{**} + 2 \le 2P\cos{\theta} \le B^{**} - 2$. Since $B^{**} = 2\alpha^* A^{**}$ (see Section~\ref{sec:lfsedge}), when $A^{**} > 4$ and $\alpha^* > 1$, there exists a range of values in which $2P\cos{\theta}$ can lie. Thus, a $\theta$ between $0$ and $\pi/2$ satisfies the inequality.
\end{proof}

I will now show that a circumcenter insertion does not add vertices too close to existing vertices. I will show that the value of $B^{**}$ should tend to infinity as $\alpha^*$ tends to 1 to ensure that the algorithm terminates with a graded mesh. The lemma below also implies that the shortest edge of a triangle is at most of length $f/B^{**}$, where $f$ is the LFS at one of the vertices. 

\begin{lemma}
\label{lemma:skinnyalpha}
Let $\bigtriangleup abc$ be a skinny triangle with circumcenter $o$. Assume that the 
lengths of all sides are greater than or equal to $\frac{f}{B^{**}}$, 
where $f$ is the LFS at one of the endpoints. 
If $\frac{B^{**}}{\alpha^*} \le B^{**} - 2$, the length of radii 
$|oa| = |ob| = |oc|$ is greater than or equal to $\frac{f}{B^{**}}$, 
where $f$ is the feature size at the $a$, $b$, $c$, or $o$.
\end{lemma}

\begin{proof}
Without loss of generality, let $ab$ be the shortest side. 
As $\bigtriangleup abc$ is skinny, the LFS-edge ratio of $oa$ (at $a$) is 
less than $\frac{B^{**}}{\alpha^*}$. If the LFS at $a$ is denoted by $f_a$, 
$f_a/|oa| \le \frac{B^{**}}{\alpha^*}$, which implies $|oa| \ge f_a/B^{**}$. 
From Lemma~\ref{lemma:diffone} above, we know that the LFS-edge ratio differs by 
at most $1$ between its endpoints. Thus, 
$\frac{f_o}{|oa|} \le \frac{B^{**}}{\alpha^*} + 1$, which implies 
$\frac{f_o}{|ob|} \le \frac{B^{**}}{\alpha^*} + 1$ because $|oa| = |ob|$, 
and $\frac{f_b}{|ob|} \le \frac{B^{**}}{\alpha^*} + 2$ 
and $\frac{f_c}{|oc|} \le \frac{B^{**}}{\alpha^*} + 2$. 
Thus, if $B^{**}/\alpha^* \le B^{**} - 2$, the lemma follows. 
\end{proof}

The following lemma tells us when the circumcenter of a 
triangle can be added even when it is not skinny. It defines the threshold
for the radius of the circumcircle for which the circumcenter
must be introduced in the facet mesh.
\begin{lemma}
\label{lemma:bigalpha}
Let $\bigtriangleup abc$ be a triangle with circumcenter $o$. 
Assume that the lengths of all sides are greater than 
$\frac{f}{B^{**}}$, where $f$ is the LFS at one of the endpoints. 
If the LFS-edge ratio of radius $oa$ at $a$ is less than or equal to $B^{**} - 2$,
the LFS-edge ratio of the radial edge at $o$, $b$ and $c$ is 
less than or equal to $B^{**}$
\end{lemma}
\begin{proof}
This is a direct consequence of Lemma~\ref{lemma:diffone}.
\end{proof}

The lemma above implies that the radius of the circumcircle
of a facet triangle is at most $f/(B^{**} - 2)$, where $f$ is
the LFS at one of the vertices. The two lemmas above imply
that the maximum radius-edge ratio of a facet triangle is
$\frac{B^{**}}{B^{**} - 2}$. Given $\alpha^*$, we can compute
the minimum $B^{**}$ to guarantee such a mesh. 

Note that the lemmas also imply that one can keep refining
the mesh if the radii of circumcircles of the triangles
are large enough relative to the LFS. 
Of course, the refinement is possible only when the potential
new vertices are inside the facet. In fact, a new vertex
should be outside the diametral circle of a subsegment
to ensure that a constrained Delaunay terahedralization 
of the PLC is possible. I will now derive the condition for 
which no vertex will be added inside the diametral circle 
of a boundary subsegment. 

\begin{lemma}
If $B^{**} < \sqrt{2}A^*$, no Steiner vertex is added inside the diametral circle of a boundary subsegment.
\end{lemma}
\begin{proof}
Suppose a Steiner vertex is added at a point $o$, and the LFS at a boundary vertex $v$ is $f$. I have shown above that $|ov| \ge f/B^{**}$. The longest boundary subsegment at $v$ may be of length $f/A^*$. Clearly, if $\sqrt{2}f/B^{**} > f/A^*$, the Steiner vertex is not inside the diametral circle of the boundary subsegment. The inequality simplifies to $B^{**} < \sqrt{2}A^*$. 
\end{proof}

Thus far, I have not addressed what would happen in the presence of small angles. In my previous paper~\cite{Sas21}, I demonstrated that boundary subsegments have empty diametral circles if they have been sufficiently refined. The following lemma from the paper makes the claim more rigorous. 

\begin{lemma}
\label{lemma:diawindow}
Consider $R = B^*/A^*$. There exists an $R>1$ below which a diametral circle of a subsegment does not contain vertices from an adjacent segment.
\end{lemma}

Note that the lemmas above help determine the minimum value of $A^*$, $B^*$ and $B^{**}$ for a given $\alpha^*$. If these values are used, I have shown my algorithm refines a facet adaptively. The length of an edge is bounded from below by $f/B^{**}$, where $f$ is the LFS at one of the endpoints of the edge. The radius of its circumcircle is bounded from above by $\alpha^* f/B^{**}$. Therefore, all triangles have a radius-edge ratio less than or equal to $\alpha^*$. Besides, the diametral spheres of all boundary edges are empty. Thus, one can construct a CDT of the domain with these added points. 

\begin{theorem}
A CDT of the domain can be constructed using the input and Steiner vertices added by the 2D LFS-based advancing-front refinement algorithm.
\end{theorem}

\begin{proof}
I ensure that diametral spheres of input subsegments are empty. See Shewchuk's paper~\cite{She98} for a proof on why empty diamtral sphere
guarantee a constrained Delaunay tetrahedralization.
\end{proof}

\subsection{The 3D Advancing-Front Refinement}
An advantage of CDT over truly Delaunay tetrahedralization is that one can place Steiner vertices inside the diametral spheres of triangular elements on input facets. An algorithm has to ensure that the radius-edge ratio of a tetrahedron within a diametral sphere is bounded from below. If not, the tetrahedron cannot be removed without refining the facet triangle because the circumcenter of the tetrahedron lies on the other side of the facet. 

\begin{figure}
\centering
\begin{tikzpicture}[scale=2]
	\draw [line width=0.25mm] (0,0) ellipse (1 and 0.3);
	\draw [line width=0.25mm, fill=none] (0,0) circle (1);
	\draw [line width=0.25mm, fill=black] ($(10:1 and 0.3)$) circle (0.05);
	\draw [line width=0.25mm, fill=black] ($(100:1 and 0.3)$) circle (0.05);
	\draw [line width=0.25mm, fill=black] ($(240:1 and 0.3)$) circle (0.05);
	\draw [line width=0.25mm] ($(240:1 and 0.3)$) -- ($(100:1 and 0.3)$) -- ($(10:1 and 0.3)$) -- ($(240:1 and 0.3)$);
	\draw [line width=0.25mm, dashed] (0,1.25) -- (0, -1.25);

	\draw [line width=0.25mm, fill=black] (0, 0.577) circle (0.05);
	\draw [line width=0.25mm, fill=black] (0, -0.577) circle (0.05);
	\draw [line width=0.35mm, dotted, fill=none] (0,-0.577) circle (1.154);

    \node [right] at ($(10:1 and 0.3)$) {a};
    \node [below right] at ($(100:1 and 0.3)$) {b};
    \node [below left] at ($(240:1 and 0.3)$) {c};
    \node [above right] at (0,0.577) {p};
    \node [right] at (0,-0.577) {o};
\end{tikzpicture}
\caption{A triangular element $\bigtriangleup abc$ on a boundary facet 
	and its diametral sphere are shown. 
	Let its radius be $r$. Let $l$ be the line
	perpendicular to the triangle and passing through its circumcenter.
	If an upper bound, $\omega^*$ on the radius-edge ratio is desired, 
	we should find a vertex $p$ such that it is at least a distance $\omega^* r$
	from $a$, $b$, and $c$. In addition, the radius-edge ratio of 
	tertrahedra $abcq$ should be at most $\omega^*$ (one cannot split this 
	tetrahedron as its circumcenter is on the other side of the facet). 
	Since the minimum possible length of the shortest
	edge of $\bigtriangleup abc$ is $r/\alpha^*$ and $\alpha^* > 1$,
	the upper bound of the radius-edge
	ratio is $\omega^* = r/|op|$. Solving for
	$\omega^*$, we get $\omega^* = 2/\sqrt{3}$, when $p$ is at a distance of 
	$r/\sqrt{3}$ from the circumcenter of the triangle. The circumcenter of 
	the tetrahedron $o$ is on the other side of the triangle, and it is 
	also at a distance of $r/\sqrt{3}$ from the circumcenter of the triangle.
	The circumsphere (dotted) is shown.
	}
\label{fig:cdtbound}
\end{figure}
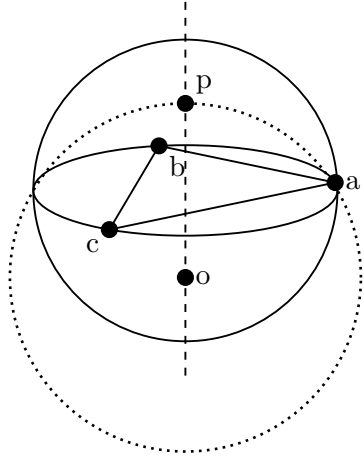

To avoid refining a facet boundary triangle, the parameter $\omega^*$ forces the algorithm to place Steiner vertices at a distance of more than $\omega^*r$ from a boundary vertex, where $r$ is the radius of the circumcircle of a triangle on an input facet.  If $\omega^*$ is close to $1$, a Steiner vertex may be placed very close to the facet boundary triangle near its circumcenter. Thus, the tetrahedron on the boundary may have a large radius-edge ratio. If $\omega^*$ is large, however, the interior tetrahedra will have a large radius-edge ratio. Thus, $\omega^*$ has to be chosen such that it minimizes the maximum of the radius-edge ratios of the interior and boundary tetrahedra. That value is $2/\sqrt{3}$. Fig.~\ref{fig:cdtbound} illustrates how I arrived at this value.

I will show below that as $\omega^*$ tends to $2/\sqrt{3}$, the value of $B^{**}$, which is inversely related to edge length, tends to infinity. Before I can analyze the algorithm, I have to define one more term. 

I will now define advancing front layers of Steiner vertices. After the input line segments are refined, consider the shortest subsegment. Let its length be $l_0$. All vertices in subsegments whose length $l$ is such that $(\omega^*)^n l_0 \le l < (\omega^*)^{n+1} l_0$ is considered to be part of the $n^{\mathrm{th}}$ layer. Since a vertex may belong to many subsegments, consider its shortest subsegment to define its layer. As the input facets are refined, again consider the shortest edge joining a new Steiner vertex with existing vertices, and assign its layer. Of course, I ignore short edges that join vertices on two input segments or facets that form a small angle. Note that the layer is assigned as soon as a vertex is inserted (not after the whole mesh is generated).

When the volume mesh is being refined, I similarly assign the layer. Let the smallest layer assigned to vertices of a skinny tetrahedron, $t$, be $l$. If a Steiner vertex, $v$, is added to split $t$, its layer assignment is at least $l+1$ because any new vertex is added at a distance of $\omega^*l$ from other vertices in the mesh. I have shown in my previous paper~\cite{Sas21} that the maximum LFS at vertex on layer $n$ is less than $l_0 (B^{**}+ \omega^* + (\omega^*)^2 + (\omega^*)^3 + ... + (\omega^*)^{n-1} + \lambda_0(\omega^*)^{n})$, where $1 \le \lambda_0 < \omega^*$. Note that the inequality holds only when we prioritize a skinny tetrahedra with shortest edges first (as I do in the algorithm).

I will now show that if $\omega^* > 2/\sqrt{3}$ and $B^{**}$ is large enough, no Steiner vertex will be introduced such that it encroaches upon a triangle on a boundary facet. 

\begin{lemma}
\label{lemma:encroachmentcdt}
If 
$$
	\left(\frac{2\alpha^*}{\omega^*\sqrt{3}}\right) \left( 1 
	+ \left(\frac{\omega^*}{\omega^* - 1}\right) \frac{1}{B^{**}}
	+ \frac{2\omega^*}{ B^{**}}
\right) < 1,
$$
no Steiner vertex will be inserted within the diametral sphere of a triangular element on an input facet such that the radius-edge ratio of the tetrahedron it forms with the triangle is less than $2/\sqrt{3}$.
\end{lemma}

\begin{proof}
	Let a Steiner vertex be introduced on layer $n \ge 1$ at $v$. The LFS at $v$ is less than $l_0(B + (\omega^*) + (\omega^*)^2 + ... + (\omega^*)^{n-1} + \lambda_0 (\omega^*)^n$, where $1 \le \lambda_0 < \omega^*$. Consider a vertex on a facet boundary triangle. This vertex is at a distance of at least $d_{\mathrm{min}} = \lambda_1 (\omega^*)^n l_0$ from $v$, where $\lambda_1 \ge \lambda_0$. Thus, the maximum LFS at the boundary vertex is $f_{\mathrm{max}} = l_0(B^{**} + (\omega^*) + (\omega^*)^2 + ... + (\omega^*)^{n-1}) + \lambda_0 l_0 (\omega^*)^n + \lambda_1 l_0 (\omega^*)^n$. The largest possible radius of the diametral sphere of a triangle adjacent to boundary vertex is
	$$r_{\mathrm{max}} = \frac{ \alpha^*f_{\mathrm{max}} }{B^{**}} = \frac{\alpha^*(l_0(B^{**} + (\omega^*) + (\omega^*)^2 + ... + (\omega^*)^{n-1}) + \lambda_0 l_0 (\omega^*)^n + \lambda_1 l_0 (\omega^*)^n)}{B^{**}}.$$
If $\frac{2}{\sqrt{3}} r_{\mathrm{max}} < d_{\mathrm{min}}$, I have proved the lemma. 
Thus, We want
$$
	\frac{2\alpha^*}{\sqrt{3}} \left(\frac{l_0(B^{**} + (\omega^*) + (\omega^*)^2 + ... + (\omega^*)^{n-1}) + \lambda_0 l_0 (\omega^*)^n + \lambda l_1 (\omega^*)^n}{B^{**}}\right) <
	\lambda_1 l_0 (\omega^*)^n.
$$
Canceling $l_0$ on both side, 
$$
	\frac{2\alpha^*}{\sqrt{3}} \left(\frac{(B^{**} + (\omega^*) + (\omega^*)^2 + ... + (\omega^*)^{n-1}) + \lambda_0 (\omega^*)^n + \lambda_1 (\omega^*)^n }{B^{**}}\right) <
	\lambda_1 (\omega^*)^n.
$$
Dividing both sides by $\lambda_1 (\omega^*)^n$ and rearranging, we get
$$
\left(\frac{2\alpha^*}{\sqrt{3}} \right) \left(
	\frac{1}{\lambda_1(\omega^*)^n} +
	\left(\frac{1}{B^{**}}\right)\left(\frac{1}{(\omega^*)^{n-1}} + \frac{1}{(\omega^*)^{n-2}} + ... \frac{1}{(\omega^*)}\right) +
	\frac{\lambda_0}{\lambda_1 B^{**}} + \frac{1}{B^{**}}
\right) < 1.
$$
Note the sum of the geometric progression in the LHS above is maximized when $n$ tends to infinity. The maximum possible value of the sum is $\frac{1}{\omega^* - 1}.$ The first term in the LHS is maximized when $n = 1$ and $\lambda_1 = 1$, and the second-to-last term is maximized when $\lambda_0=\lambda_1$. Thus, the lemma is proved when
$$
	\left(\frac{2\alpha^*}{\omega^*\sqrt{3}}\right) \left( 1 
	+ \left(\frac{\omega^*}{\omega^* - 1}\right) \frac{1}{B^{**}}
	+ \frac{2\omega^*}{ B^{**}}
\right) < 1.
$$
\end{proof}

Note that the inequality above can be satisfied by choosing a large enough $B^{**}$ when $\alpha^* > 1$ and $\omega^* > 2\alpha^*/\sqrt{3}$, and the minimum value of $B^{**}$ tends to infinity as the value of $\omega^*$ tends to $2/\sqrt{3}$. 

Also, no vertex is placed ``near'' a facet vertex, i.e., there is a sphere around a facet vertex where no other vertex is allowed.  I call this the forbidden region. A ``permitted region'' is also present inside a diametral sphere of a facet where Steiner vertices are allowed. Consider the circumcenter of a tetrahedron formed by a vertex in the permitted region and vertices of the facet triangle. The circumcenter is in the forbidden region on the other side of the facet. Since the radius-edge ratio of the tetrahedron is considered to be good, there is no need to add the circumcenter. 

I have shown above that a tetrahedron formed by a vertex and a facet's triangle has a desirable radius-edge ratio. I have to also show that no other poor-quality tetrahedra (that do not contain the facet triangle) is present inside the diametral sphere of the triangle. 

\begin{lemma}
A tetrahedron cannot be of poor quality if it is inside a facet's triangle's diametral sphere and its circumcenter is on the other side of the facet. 
\end{lemma}
\begin{proof}
Let us assume such a tetrahedron exists. Consider the diametral sphere of the facet triangle. On either side of the facet, there are two regions in the diametral hemisphere. Per the lemma above, vertices are permitted in one of the regions, and vertices are forbidden in the other region. Besides, since the circumcenter of the tetrahedron is a potential vertex, the circumcenter cannot be in the forbidden region on either side of the facet if the lemma above holds. But if the circumcenter is in the permitted region on the other side of the facet, no part of the circumsphere is in the permitted region on current side of the facet. Thus, the poor-quality tetrahedra cannot exist. 
\end{proof}

I will now show that the algorithm terminates with a graded mesh. For vertices on input line segments and input facets, I have already shown that the LFS-edge ratio is at most $B^*$ and $B^{**}$, respectively. I will derive the LFS-edge ratio for the vertices in the interior. 

\begin{theorem}
\label{lemma:gradedcdt}
The algorithm terminates with a graded mesh if $\omega^* > 2/\sqrt{3}$.
\end{theorem}

\begin{proof}
Consider a Steiner vertex, $v$, inserted in the $n^{\mathrm{th}}$ layer. When $v$ is inserted, the length of the shortest edge adjacent to $v$ is at least $l_{\mathrm{min}} = l_0 \lambda (\omega^*)^{n}$, where $l_0$ is the shortest edge among the subsegments after the input segment refinement and $1 \le \lambda < \omega^*$. The LFS at $v$ is at most $f_{\mathrm{max}} = l_0(B^{**} + \omega^* + (\omega^*)^2 + ... + (\omega^*)^{n - 1} + \lambda (\omega^*)^n)$. Let $t$ be the skinny tetrahedron that was split to insert $v$. Because I prioritize skinny tetrahedron with the shortest edge first, and because I place vertices in an advancing-front manner, the length of the shortest edge of $t$ is at least 
$$l_{\mathrm{parent}} = \frac{l_{\mathrm{min}}}{\Gamma \lambda \omega^*} = \frac{l_0 \lambda(\omega^*)^{n}}{\Gamma \omega^*} = \frac{l_0 \lambda(\omega^*)^{n-1}}{\Gamma}.$$ 
Any vertex inserted after this tetrahedron is split will be at least a distance $l_{\mathrm{parent}} \omega^* = \frac{l_0 \lambda(\omega^*)^{n-1}} {\Gamma} \omega^* = \frac{l_0 \lambda(\omega^*)^{n}} {\Gamma}$ from other vertices in the mesh. There the maximum LFS-edge ratio is
$$
\frac{l_0(B^{**} + \omega^* + (\omega^*)^2 + ... + (\omega^*)^{n} + \lambda(\omega^*)^{n})}
{\frac{l_0 \lambda(\omega^*)^{n}}{\Gamma}}, 
$$
which simplifies to
$$
\left(\frac{\Gamma}{\lambda}\right)\left(\frac{ B^{**}}{(\omega^*)^{n}} + 
	\frac{1}{(\omega^*)^n} + ... +
	\frac{1}{(\omega^*)^2} + \frac{1}{(\omega^*)} + 1 + 
	\lambda\right).$$
As $\lambda \ge 1$, the expression above is less than
$$
\Gamma \left(B^{**} + \frac{\omega^*}{\omega^* - 1} + 1\right).
$$
Since the LFS-edge ratio is bounded, the algorithm terminates with a graded mesh. 
\end{proof}

\section{Extension to Truly Delaunay Mesh Refinement}
The Steiner vertices placed in the CDT algorithm above may not recover input facets if their true Delaunay tetrahedralization, as opposed to CDT, is constructed. A sufficient condition that recovers a facet is to ensure that the diametral spheres of all triangular elements on the facet are empty. This condition is also called Gabriel's condition.  For nonadjacent features, it is easy to ensure that Gabriel's condition is satisfied by sufficiently refining the mesh. For adjacent feature with a dihedral angle greater than $\pi/2$, Gabriel's condition is satisfied for all triangles in the facets without additional measures. When the dihedral angle between two facets or a facet and a line segment is small, however, extra steps are necessary to ensure that the vertices on the facets and line segments are themselves not inside diametral spheres of triangles on adjacent facets. Note that the satisfaction of Gabriel's condition is sufficient for boundary recovery, but it not necessary. In this section, I will describe these extra steps in detail. 

\subsection{Mirroring}

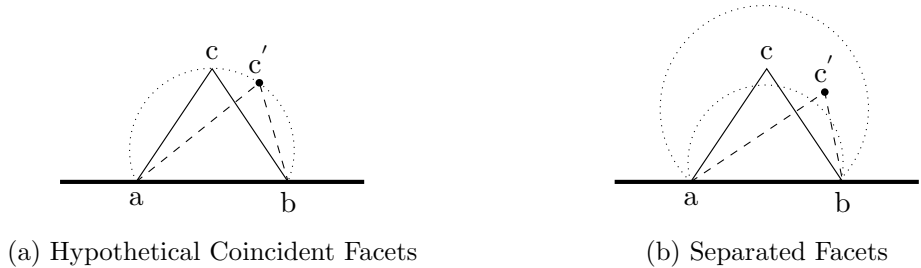
\begin{figure}
\centering
\begin{subfigure}[b]{0.45\textwidth}
\centering
\begin{tikzpicture}[scale=1]
\draw [line width=0.55mm] (-2,0) -- (2,0);
\draw [line width=0.15mm] (-1,0) -- (0,1.5) -- (1,0);
	\draw [dotted] (-1,0) arc (203:-23:1.083);
    \node [below] at (-1,0) {a};
    \node [below] at (1,0) {b};
    \node [above] at (0,1.5) {c};
\coordinate (center) at (0,0.427);
	\coordinate (cdash) at ($ (0,0.427)+(55:1.083) $) ;
	\fill[black] (cdash) circle[radius=0.05];
	\node [above] at (cdash) {c$^{'}$};
	\draw [dashed] (-1,0) -- (cdash) -- (1,0);
\end{tikzpicture}
\caption{Hypothetical Coincident Facets}
\end{subfigure}
\centering
\begin{subfigure}[b]{0.45\textwidth}
\centering
\begin{tikzpicture}[scale=1]
\draw [line width=0.55mm] (-2,0) -- (2,0);
\draw [line width=0.15mm] (-1,0) -- (0,1.5) -- (1,0);
	\draw [dotted] (-1,0) arc (195:-15:1.02);
	\draw [dotted] (-1,0) arc (225:-45:1.37);
    \node [below] at (-1,0) {a};
    \node [below] at (1,0) {b};
    \node [above] at (0,1.5) {c};
\coordinate (center) at (0,0.427);
\coordinate (cdash) at ($ (0,0.427)+(45:1.08) $) ;
\fill[black] (cdash) circle[radius=0.05];
	\node [above] at (cdash) {c$^{'}$};
	\draw [dashed] (-1,0) -- (cdash) -- (1,0);
\end{tikzpicture}
\caption{Separated Facets}
\end{subfigure}
\caption{
When two facets sharing an input segment (thick line 
segment) are separated by 
a small dihedral angle, a vertex on one facet may be 
inside the diametral sphere of a triangle on the other 
facet. In order to avoid this situation, i.e., the violation 
of Gabriel's condition, I mirror the 
triangle on one facet on to all other facets that
share the input segment. 
(a) A hypothetical case where the facets are coincident.
The free vertex $c^{'}$ has to lie on the the circumcircle
of triangle $\bigtriangleup abc$.
(b) A case where the facets are separated by a small angle.
A free vertex may be placed in a window around the 
mirrored circumcircle. 
}
\label{fig:mirror}
\end{figure}

Consider two facets that meet at a small angle $\phi$ on a line segment $l$. Consider a region somewhere in the middle of the line segment, where a short subsegment, $l_{ab}$, is present that joins two vertices $a$ and $b$. Consider two triangles containing $l_{ab}$ on the two facets. Let those triangles be $\bigtriangleup abc$ and $\bigtriangleup abc^{'}$, where vertices $c$ and $c^{'}$ are on the adjacent facets. If the facets are coincident ($\phi = 0$), $c^{'}$ has to be on the circumcircle of $\bigtriangleup abc$. Otherwise, either $c^{'}$ is inside the diametral sphere of $\bigtriangleup abc$ or $c$ is inside the diametral sphere of $\bigtriangleup abc^{'}$ (see Fig.~\ref{fig:mirror}). As we increase the $\phi$ to larger values, there is a larger window where $c{'}$ may be placed so that Gabriel's condition is satisfied for both triangles. This observation leads me to construct meshes ``near'' the line of intersection of two facets such that the meshes on the facets mirror each other. 

I will call the region where the meshes are mirrored the cylindrical boundary layer. I will show later that the radius of the ``cylinder'' varies along the line of intersection, but the region is isotopic to a cylinder. If multiple facets are present on a line with small angles between them, the surface meshes on the facets have to be mirrored. Besides, the meshes should also be refined proportionally to the local feature size. Both these objectives can be achieved by refining the input line segment finely and using the advancing front technique described in Section 3.2. Finely refining the input line segment ensures that other nonadjacent features are far away relative to the length of subsegments. Also, the LFS on the vertices of triangles near the line of intersection $l$ will nearly be identical. If $B^{*}$ and $B^{**}$ are large enough, the triangles outside the cylindrical boundary layer will be so small that their diametral spheres will not intersect adjacent or nonadjacent features. Later in this section, I will formally prove the statements above. 

\subsection{Split on Concentric Spheres}

\begin{figure}
\centering
\begin{subfigure}[b]{0.45\textwidth}
\begin{tikzpicture}[scale=2]
    \draw [line width=0.55mm] (0,0) -- (30:1) -- (30:2);
    \draw [line width=0.55mm] (0,0) -- (-30:1) -- (-30:2);
    \draw [line width=0.15mm] (0,0) -- (-25:1) -- (-25:2);
    \draw [line width=0.15mm] (0,0) -- (-20:1) -- (-20:2);
    \draw [line width=0.15mm] (0,0) -- (-10:1) -- (-10:2);
    \draw [line width=0.15mm] (0,0) -- (0:1) -- (0:2);
    \draw [line width=0.15mm] (0,0) -- (10:1) -- (10:2);
    \draw [line width=0.15mm] (0,0) -- (17:1) -- (17:2);
    \draw [line width=0.15mm] (0,0) -- (24:1) -- (24:2);

    \draw [line width=0.15mm] (30:1) -- (24:2);
    \draw [line width=0.15mm] (24:1) -- (17:2);
    \draw [line width=0.15mm] (17:1) -- (10:2);
    \draw [line width=0.15mm] (10:1) -- (0:2);
    \draw [line width=0.15mm] (0:1) -- (-10:2);
    \draw [line width=0.15mm] (-10:1) -- (-20:2);
    \draw [line width=0.15mm] (-20:1) -- (-25:2);
    \draw [line width=0.15mm] (-25:1) -- (-30:2);

    \draw [line width=0.15mm] (30:1) -- (24:1) -- (17:1) -- (10:1) -- (0:1) -- (-10:1) -- (-20:1) -- (-30:1);
    \draw [line width=0.15mm] (30:2) -- (24:2) -- (17:2) -- (10:2) -- (0:2) -- (-10:2) -- (-20:2) -- (-30:2);
\end{tikzpicture}
\caption{The Modified SOS Technique}
\end{subfigure}
\centering
\begin{subfigure}[b]{0.45\textwidth}
\begin{tikzpicture}[scale=2]
    \draw [line width=0.55mm] (0,0) -- (30:2);
    \draw [line width=0.55mm] (0,0) -- (-30:2.24);
    \draw [line width=0.15mm] (0,0) -- (20:2.04);
    \draw [line width=0.15mm] (0,0) -- (10:2.08);
    \draw [line width=0.15mm] (0,0) -- (0:2.12);
    \draw [line width=0.15mm] (0,0) -- (-10:2.16);
    \draw [line width=0.15mm] (0,0) -- (-20:2.20);
    \draw[dotted] ([shift=(30:2)]0,0) arc (30:-30:2);
    \draw[dotted] ([shift=(30:2.24)]0,0) arc (30:-30:2.24);
\end{tikzpicture}
\caption{Radial Linear Interpolation}
\end{subfigure}
\caption{
The splitting of a sector on an ``apex'' vertex. I use a modified
SOS technique.
The thick edges are part of the input PLC. 
(a) The eventual result of my algorithm after all the splitting.
(b) Since the lengths of the subsegments adjacent to $o$ on the PLC
are not identical,
some linear interpolation is required. The difference in their lengths
is exaggerated to show it prominently.
}
\label{fig:socs}
\end{figure}
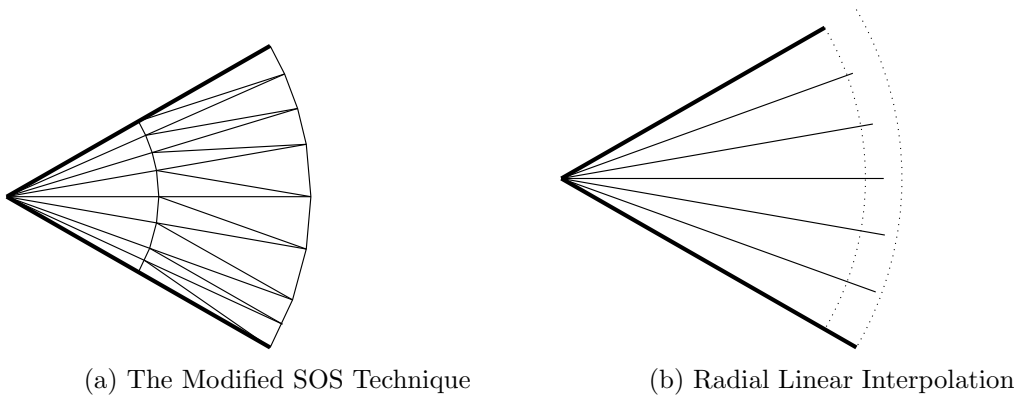

Consider a vertex from which several input facets and input line segments ``emanate''. Such a vertex can be the apex of a pyramidal structure, for instance. I will call this vertex the apex vertex. For such domains, prior researchers have used the split-on-the-sphere (SOS) technique~\cite{CVY02} to ensure that the facets are recovered by Delaunay tetrahedralization. In their techniques, they add vertices only on small concentric spheres that is centered at the vertex with a small radius proportional to the LFS at the vertex. The intersection of the sphere and a facet is a sector of a circle on the facet. On the circumference of the sector, their algorithms continually add Steiner vertices such that the sectors are refined into multiple triangles until the boundary is recovered by Delaunay tetrahedralization. I extend the SOS technique in this paper.

In my previous paper~\cite{Sas21}, I refined input segments that formed a small angle until Gabriel's condition was satisfied, i.e., the diametral circles of subsegments (after the 1D ODE-based refinement) were empty. As a result of extensively refining the input segments, the value of $B^*$ would end up being bigger for smaller angles. Consequently, the value of $R = B^*/A^*$ ended up being arbitrarily close to 1, which results in input subsegments being of almost nearly identical lengths near a vertex with two or more segments forming a small angle. This segment refinement is reminiscent of Ruppert's technique~\cite{R93,R95} of splitting on concentric shells when dealing with small angles in 2D PSLGs. 

In my technique to deal with small angles in 3D, I use a modified SOS technique. I refine the input segments sufficiently (as a function of the angles between facets) until $R = B^*/A^*$ is close to $1$. The refinement should be sufficient enough that other nonadjacent features in the input are far away relative to the lengths of the subsegments near the apex vertex $o$. The lengths of adjacent subsegments differ by a factor of at most $R$ because they share a vertex. This refinement results in subsegments of nearly identical lengths emanating from the apex vertex. Vertices will be placed in concentric ``spheres'' around the apex vertex. I then refine the sectors on the facet sufficiently so that the facets are recovered after Gabriel's condition is satisfied. Fig.~\ref{fig:socs}(a) provides an illustration of how a set of concentric shells will eventually be split and meshed. As illustrated, I refine all the sectors present in the ``spherical'' boundary layer around the apex vertex. The value of $B^{**}$ is chosen such that the triangles outside the spherical (and cylindrical, too) boundary layer are so small that their diametral sphere does not intersect with adjacent or nonadjacent features. 

I still have not answered the following question: when multiple facets and segments are emanating from an apex vertex, where and how should one refine the sectors? To refine a sector, I use a technique similar to the one I use for input segment refinement. For input segment refinement, I consider the LFS function along a segment and use the solution of piecewise-smooth ODE to decide where to split an input segment. I do the same for a sector, but instead of the LFS function, I use the angular feature size (AFS) function I describe below. This modification ensures that I adaptively refine sectors near regions where facets and line segments meet at small angles.
 
\subsubsection{Angular Feature Size}

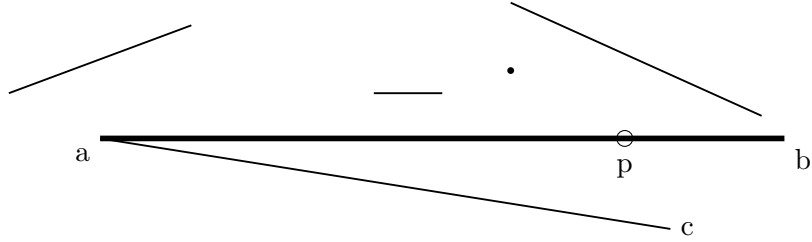
\begin{figure}
\centering
\begin{tikzpicture}[scale=3]
	\draw [line width=0.75mm] (0,0) -- (3,0);
	\draw [line width=0.25mm] (0,0) -- (2.5,-0.4);
	\draw [line width=0.25mm] (-0.4,0.2) -- (0.4,0.5);
	\draw [line width=0.25mm] (1.2,0.2) -- (1.5,0.2);
	\draw [line width=0.25mm] (1.8,0.6) -- (2.9,0.1);
	\fill[black] (1.8, 0.3) circle[radius=0.015];
	\draw (2.3, 0.0) circle[radius=0.035];
	    \node [below left] at (0,0) {a};
	    \node [below right] at (3,0) {b};
	    \node [below] at (2.3,-0.05) {p};
	    \node [right] at (2.5,-0.4) {c};
\end{tikzpicture}
	\caption{Consider an ``apex'' vertex $o$ (not shown) and all 
	its adjacent features. Consider a small sphere 
	that intersects all the features. This diagram
	is the Mercator projection of those features. Note that
	a facet adjacent to $o$ is a line segment here, and 
	a line segment adjacent to $o$ is a vertex. The AFS 
	at a point $p$ on a facet $oab$ ($p$, $a$, and $b$ are on 
	the sphere) is analogous to the LFS on the Mercator projection.
	The facet represented by $ac$ intersects the facet represented by
	$ab$ at a line. When computing the AFS of ray $op$ on the facet 
	represented by $ab$, I ignore the facet
	represented by $ac$, just as in computation of the LFS.}
	\label{fig:mercator}
\end{figure}

Consider an apex vertex, $o$, and an adjacent facet, $f$. 
Informally, the angular feature size (AFS) is the LFS on a small sphere with center $o$ that intersects all features adjacent to $o$, and the LFS is normalized with respect to the radius of the sphere (see Fig.~\ref{fig:mercator}). 
Formally, consider all other adjacent facets (adjacent to $f$) and input segments (adjacent to $o$), but exclude the facet that intersects $f$ on a line. Consider a sphere, however small, centered at $o$ that intersects all these features. I will define the AFS on this sphere, $s$. Let $oa$ and $ob$ be the input segments that define the border of the facet $f$. The AFS is a function of angle $\theta$ that a vector makes with $oa$ as it rotates from $oa$ to $ob$. Let that vector be $\overrightarrow{op}$, which intersects $s$ at $x$. The non-normalized AFS at $\theta$ is simply the distance from $x$ to the nearest feature on $s$ measured on $s$. Note the nearest feature, $y$, should not be on the facet $f$. Also, note that the distance is measured on a greater circle joining the $x$ and $y$. The normalized AFS is equal to the non-normalized AFS divided by the radius of $s$. The normalized AFS is essentially $\sin{\angle xoy}$. Thus, the AFS is smaller when the angle between facets is smaller. It is well defined even when multiple facets and line segments are emanating from $o$. Fig.~\ref{fig:mercator} shows the relationship between LFS and AFS by considering the Mercator projection of the features on $s$. 

\subsubsection{Sector Refinement}

After the input segments are refined, consider a facet adjacent to an apex vertex $o$. Adjacent to $o$, two subsegments form the border of the facet. Let those subsegments be $oa$ and $ob$. Since $R$ is close to 1, the lengths of $oa$ and $ob$ are nearly identical. Consider the vector $\overrightarrow{oa}$. As one rotates the vector to $\overrightarrow{ob}$, one can linearly interpolate its magnitude. I refine this ``sector'' with linearly varying radius (see Fig.~\ref{fig:socs}(b)). The SOS refinement is identical to how I refine input line segments. This refinement can be independently carried out on every apex vertex. Just as for the LFS-edge ratio, this refinement has an analogous AFS-angle ratio, whose values are restricted to be between some $A^{'}$ and $B^{'}$. As we refine more, the ratio $R^{'} = B^{'}/A^{'}$ tends to 1. The ratio $R^{'}$ also dictates the maximum ratio of the magnitude of adjacent angles of a refined sector (just as the lengths of adjacent subsegments have a ratio less than $R = B^*/A^*$).
Note that the ODE needs to be solved numerically, and since the derivative 
of the solution of the ODE never vanishes (it is a monotonically increasing 
function), numerical techniques should be able to solve it accurately. 
Also note that tetrahedra that contain these sector triangles have
to be evaluated using the modified definition of a skinny tetrahedron
because one of their sides is very short due to the small angles around
the apex vertex.
\subsection{Interaction Between Boundary Layers}

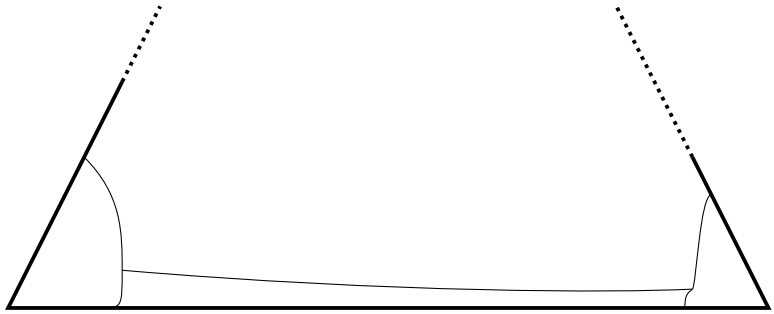
\begin{figure}
	\centering
\begin{tikzpicture}
	\draw [line width=0.50mm] (1.5,3) -- (0,0) -- (10,0) -- (9,2);
	\draw [line width=0.50mm, dotted] (1.5,3) -- (2,4);
	\draw [line width=0.50mm, dotted] (9,2) -- (8,4);
\draw (1,2) .. controls (1.5,1.5) and (1.5,1) .. (1.5,0.5);
\draw (1.5,0.5) .. controls (1.5,0) .. (1.2,0);
\draw (9.25,1.5) .. controls (9.1,1.5) and (9.05,0.3) .. (9,0.25);
\draw (9,0.25) .. controls (8.95,0.2) and (8.9, 0.2).. (8.9,0);
\draw (1.5, 0.5) .. controls (5,0.2) and (8, 0.2).. (9,0.25);
\end{tikzpicture}
	\caption{An illustration of spherical and cylindrical boundary layers on a facet.
	The thick lines are the boundaries of the facet. The curved lines represent the
	boundary layers. Note the radius of the boundary layers
	is a function of the AFS, LFS, and 
	the angles between input facets and line segments.}
	\label{fig:bl}
\end{figure}

One can define cylindrical boundary layers to be small such that they do not interact among themselves. This observation is also true for spherical boundary layers. A natural question about the interaction between cylindrical and spherical boundary layer arises: Do they interact favorably so that Gabriel's condition is satisfied where they interact? 

Thankfully, only the outer layer of the spherical boundary layer interacts with one end of the cylindrical boundary layer. If the cylindrical boundary later is refined sufficiently, the triangular meshes in adjacent facets are nearly mirrored. The AFS within the intersection of cylindrical and spherical boundary layers does not vary much if $R^{'}$ is sufficiently close to $1$. As a result, the triangles in the cylindrical and spherical boundary layers are almost identical, and they satisfy Gabriel's condition. I will prove this statement formally in the next section. See Fig.~\ref{fig:bl} for an illustration of the spherical and cylindrical boundary layers. 

\subsection{The Algorithm}
The following algorithm outputs a graded truly Delaunay mesh that respects Gabriel's condition. For the algorithm, I pick a large enough $B^{**}$ and $B^{'}$. I will show in the next section how their values depend on $\omega^*$, LFS, and AFS. 

\begin{enumerate}
\item Refine 1D input segments.
\item Use the SOS technique to split sectors adjacent to apex vertices.
\item For all input line segments, pick any facet adjacent to an input segment and use the advancing-front algorithm to refine the mesh in the boundary layer.
\item Mirror the mesh on other facets in the boundary layer. 
\item Refine the mesh on the rest of the facet (outside the boundary layer) by either circumcenter insertion or the advancing-front algorithm.
\item Tetrahedralize the vertices and refine the tetrahedral mesh until all elements have a desirable radius-edge ratio.
\end{enumerate}

\section{An Analysis of the Truly Delaunay Mesh Refinement Algorithm}
In my analysis, I will first consider facet meshing. I first show that it is possible to construct cylindrical and spherical boundary layers such that Gabriel's condition is satisfied outside the boundary layer. I have already argued why mirroring satisfies Gabriel's condition insider a cylindrical boundary layer. I will then derive conditions that ensure that Gabriel's condition is satisfied inside the spherical boundary layer. After these steps, I will consider volume meshing and show that my algorithm generates well-graded truly Delaunay tetrahedral meshes. 

The following lemma shows that Gabriel's condition is satisfied by the facet mesh outside of the cylindrical boundary layer.

\begin{lemma}
\label{lemma:cylblout}
Let the angle between two adjacent facets be $\phi$ and the LFS at a point $o$ on their line of intersection be $F$. Let the radius of the boundary layer at $o$ be $d$. If $(F + d)/B^{**} < d \sin{\phi}$, Gabriel's condition is satisfied outside the boundary layer. Note the condition has to be satisfied for all points on the line of intersection. 
\end{lemma}

\begin{proof}
The distance between the two facets outside the boundary layer is at least $d \sin{\phi}$. The radius of the diametral sphere of a triangle, all of whose vertices are outside the boundary layer, is at most $(F + d)/B^{**}$. Thus if $(F + d)/B^{**} < d \sin{\phi}$, Gabriel condition is satisfied outside the boundary layer. 
\end{proof}

The lemma above shows that for a choice of $d$, it is possible to compute $B^{**}$ for which Gabriel's condition is satisfied inside (due to mirroring) and outside the boundary layer. Thus, we should pick $d$ such that the cylindrical boundary layers do not interact. 

We also want $d$ to be small enough that mirrored triangles also satisfy the LFS-edge ratio condition. If $d$ is small, the LFS will not vary much in the boundary layer, so the LFS-edge ratio will be preserved. We should actually choose $d$ as a fraction of the LFS. I will prove that formally below.  

\begin{lemma}
\label{lemma:mirrorratio}
The LFS-edge ratio condition is satisfied by mirrored triangles. 
\end{lemma}

\begin{proof}
Consider a triangle constructed using the advancing-front algorithm on a facet in the boundary layer.  Let the LFS at one of its vertices $v$ be $F$. The LFS at a vertex of a mirrored triangle is at least $F-2d$. Let the length of an edge adjacent to $v$ be $l$. When I construct the triangle, I will ensure that it satisfies the LFS-edge ratio such that $F/l \le \gamma B^{**}$, where $\gamma > 1$ is some constant that I will determine. 
For the mirrored triangles to satisfy the LFS-edge condition, we want $(F - 2d) / l \le B^{**}$.
The LHS is maximized when $l$ is minimized, i.e., $l = F/(\gamma B^{**})$.
Thus, we want
$$
\left(\frac{\gamma B^{**}}{F}\right)(F - 2d) \le B^{**},
$$
which implies
$$
\gamma\left(1 - \frac{2d}{F}\right) \le 1.
$$
When we mirror a facet mesh in the boundary later, we pick the
radius of the boundary layer to be a fraction of the local 
feature size and compute $\gamma > 1$ based on the relationship 
above. This choice results in a larger mesh (with smaller elements) on some
facets (because $B^{**}$ increases by a factor of $\gamma$).
\end{proof}

From the two lemmas above, the algorithm becomes clearer. Based on 
the LFS at a point on the line of intersection of two facets and
the dihedral angle $\phi$ between them, I pick the radius 
of the boundary layer $d$ to be a fixed fraction of the LFS. 
This fraction dictates the minimum value of $B^{**}$ and 
the value of $\gamma$. The smaller the fraction, 
the smaller the $\gamma$. Therefore, in the algorithm, I refine a 
facet with a slightly larger $B^{**}$ 
and mirror the mesh on adjacent facets inside the cylindrical
boundary layer. 

The following lemma states that Gabriel's condition is satisfied
outside of the spherical boundary layer. The reason is that adjacent
and nonadjacent facets are separated by sufficient distance relative
to the radius of the diametral sphere of those triangles. This
reasoning is the same as for the one for Lemma~\ref{lemma:cylblout}.
\begin{lemma}
Consider an apex vertex $o$. Consider a vertex $p$ at a distance $d$ on a facet. The vertex may be an input vertex, one added by the SOS technique, or one added during meshing the facet. Among all vertices of a triangle $t$ adjacent to $p$, assume AFS at ray $op$ is the smallest. From $p$, consider a point $q$ nearest to $p$ on an adjacent input line segment or an input facet.  Let the $\angle poq$ be $\phi$ and the LFS at a point $o$ be $F$. If $(F + d)/B^{**} < d \sin{\phi}$, Gabriel's condition is not violated by $t$. 
\end{lemma}
\begin{proof}
The proof is identical to that of Lemma~\ref{lemma:cylblout}. 
\end{proof}

\begin{lemma}
The number of concentric spheres in the boundary layer is bounded from above by a constant.
\end{lemma}

\begin{proof}
In the lemma above, I have shown that $(f + d)/B^* < d \sin{\phi}$ outside the boundary layer. Thus, the radius $d$ of the boundary layer is at most $f / (B^*\sin{\phi} - 1)$. The least possible LFS in the boundary layer is $f - d$, so the least possible length of a subsegment is $(f - d)/B^*$. Therefore, the number of such segments is at most $d B^* / (f - d)$. The maximum number of subsegments is obtained when $d$ is maximized, which is $f / (B^*\sin{\phi} - 1)$. Substituting for $f$ and simplifying the equation, the maximum number of subsegment (also the maximum number of concentric spheres) is $B^*/(B^*\sin{\phi} - 2)$.
\end{proof}

I will now describe the condition under which the SOS technique eventually recovers boundary facets near apex vertices after sufficient refinement. Suppose the algorithm has refined input facets and line segments around an apex vertex such that all vertices adjacent to the apex vertex are exactly on a sphere. Consider a facet whose sector has to be recovered. If the sector is continuously split, the triangles on the sector tend to become almost a straight line, and their diametral spheres tend to be inside the sphere around the apex vertex. Since all other vertices are on the sphere, every triangle is recovered when it is sufficiently split. 

In our case, the vertices are not exactly on the sphere, but they are almost on a sphere. In that case, $B^*$ has to be large enough (so that $R = B^*/A^*$ is close enough to 1) such that it is still possible to use the SOS algorithm to satisfy Gabriel's condition. I will show below that the value of $R$ is a function of the AFS around the vertex.

\begin{figure}
\centering
\begin{tikzpicture}[scale=2]
	\draw[line width=0.25mm] (1,0) arc (0:180:0.5) -- (1,0);
	\draw[line width=0.25mm, dotted] (-20:1) arc (-20:200:1);
	\draw[line width=0.25mm] (0,0) -- (40:1.2);
	\draw[line width=0.25mm] (0.2, 0) arc (0:40:0.2);
    \node [below] at (0,0) {o};
    \node [right] at (1,0) {a};
	\node [below] at (40:0.745) {p};
	\node [right] at (30:0.2) {$\phi$};
\end{tikzpicture}
\caption{
	The vertex $o$ is an ``apex'' vertex from which multiple facets
	and line segments emanate. The line segment $oa$ is an edge on a facet
	and an edge of $\bigtriangleup oab$, which is one of the triangles 
	obtained after the SOS technique is applied. Assume $\bigtriangleup oab$ 
	is small enough. Its diametral hemisphere is shown as the semicircle. The
	ray $op$ is on an input segment. The dotted circle represents the sphere
	with center $o$ and radius $|oa|$. Note that $|op| = |oa|\sin{\phi}$. If
	$R < 1/\sin{\phi}$, a vertex on ray $op$ will not be added inside the
	diametral hemisphere. 
	}
\label{fig:socs_rdash}
\end{figure}
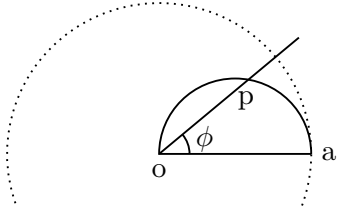

\begin{lemma}
\label{lemma:smallR}
For a small enough $R = B^*/A^*$, there exists a minimum $B{'}$ for which Gabriel's condition is satisfied within a spherical boundary layer, and that value is a function of the AFS.
\end{lemma}

\begin{proof}
Consider the smallest of the concentric spheres around the apex vertex $o$. Consider a facet sector that is being refined. In the limiting case (when the value of $B^{'}$ is very high), a triangle $t$ on the sector is almost a line. Let the length of a side of the triangle adjacent to $o$ be $1$ unit. On some other facet or line segment adjacent to $o$, a vertex is at least a distance $1/R$ from $o$ (because lengths of adjacent subsegments differ by a factor of at most R as they share a vertex). As we see in Fig.~\ref{fig:socs_rdash}, if $1/R > \cos{\phi}$, only then is that vertex outside the diametral sphere of $t$. Thus, when $B^*$ is greater than a certain threshold (depending on $\phi$), and $R$ is close enough to $1$, there exists some $B^{'}$ for which Gabriel's condition is satisfied. 

For the second smallest concentric sphere, we want $1/R^2 > \cos{\phi}$. For the $n^{\mathrm{th}}$ sphere, we want $1/R^n > \cos{\phi}$. I have shown that $n$ is a constant in the lemma above. Thus, if $R$ is close enough $1$, Gabriel's condition is satisfied by the SOS technique.   
\end{proof}

I will prove below that with enough refinement, Gabriel's condition is satisfied everywhere.

\begin{lemma}
When $R = B^*/A^*$ and $R^{'} = B^{'}/A^{'}$ are close enough to $1$, Gabriel's condition is valid everywhere. 
\end{lemma}

\begin{proof}
It is easy to show that when $B^{**}$ is greater than 2, the diametral spheres on facet triangles are so small that they do not intersect nonadjacent features. For adjacent features, I have already shown in the lemmas above that Gabriel's condition is satisfied strictly inside or outside the boundary layers. Consider the interface of spherical and cylindrical boundary layers. The vertices and triangles on the last concentric sphere are the interface layer. When $R$ and $R^{'}$ are both $1$ (hypothetically, after infinite refinement), the adjacent line segments (due to the SOS technique) have equal length. In this case, the interface triangles on adjacent facets mirror each other exactly. The exact mirroring is necessary only when the angle between the two facets vanishes. Since, there is a finite angle $\phi > 0$ between any two facets, $R$ and $R^{'}$ may be larger than 1. Depending on $\phi$, there exists some $R > 1$ and $R^{'} > 1$ such that Gabriel's condition is satisfied everywhere.
\end{proof}

In the above lemmas, I have shown that we can compute the minimum $B^*$ and $B^{**}$ such that Gabriel's condition is satisfied everywhere. Their values are some function of angles between facets and line segments. These values are also dictated by the radii of the boundary layers. Thus, $B^*$ and $B^{**}$ have to be chosen to be large enough that all these conditions are satisfied. I have also shown that there exists an $A^{'}$ and a $B^{'}$ such that Gabriel's condition is satisfied everywhere. I have not provided an explicit equation to compute those values because it is hard to derive the values for the interface of the cylindrical and spherical boundary layer. In my algorithm, I suggest that we keep refining the sectors at an apex vertex using the SOS technique until Gabriel's condition is satisfied. 

Finally, I prove that if $\omega^* > \sqrt{2}$, my technique does not introduce vertices inside a boundary triangular element's diametral sphere and that it outputs a graded mesh.

\begin{lemma}
\label{lemma:encroachmenttdt}
If 
$$
	\left(\frac{\sqrt{2}\alpha^*}{\omega^*}\right) \left( \gamma 
	+ \left(\frac{\omega^*}{\omega^* - 1}\right) \frac{1}{B^{**}}
	+ \frac{2\omega^*}{ B^{**}}
\right) < 1,
$$
no Steiner vertex will be inserted within the diametral sphere of a triangular element on an input facet.
\end{lemma} 

\begin{proof}
The proof is similar to the one for Lemma~\ref{lemma:encroachmentcdt} with $2/\sqrt{3}$ being replaced with $\sqrt{2}$. 
Note that one should also account for $\gamma$ from Lemma~\ref{lemma:mirrorratio}. It is 
possible\footnote{Informally, consider an isosceles $\bigtriangleup abc$, where $ab = ab$ and $\angle{bac} = \pi$. The radius
of the diametral circle (circumcircle) is infinity. As $\angle{bac}$ tends to 0, the radius also monotonically reduces. This
argument may be extended to the SOS refinement technique.} 	
to show that the
radii of the diametral spheres of triangles due to the SOS technique is smaller than the radii
of the diametral spheres of triangles with LFS-constrained edge lengths. Thus, all the steps
in the proof of Lemma~\ref{lemma:encroachmentcdt} apply here.
\end{proof}

\begin{theorem}
\label{lemma:gradedtdt}
The algorithm terminates with a graded mesh if $\omega^* > \sqrt{2}$.
\end{theorem}

\begin{proof}
The proof is similar to the one for Theorem~\ref{lemma:gradedcdt}.
\end{proof}

Putting everything together, the input to the algorithm is the PLC and $\omega^*$. 
The value of $\omega^*$ dictates the maximum value of $\gamma$, which dictates
the maximum value of the fraction that decides the radii of the boundary layers.
As long as the fraction is less than $1/2$ and the spherical boundary layer
is large enough, cylindrical boundary layers will not interact. 
A smaller $\omega^{*}$ (and as a result, a lower $\gamma$) 
also increases the lower bound on $B^{**}$ (see Lemma~\ref{lemma:encroachmenttdt}). 
A lower bound on $B^{*}$ (and thus, $B^{**}$) is also provided by the 
dihedral angles between 
any two facets adjacent to an apex vertex due to Lemma~\ref{lemma:smallR}. The
value of $B^{**}$ has to be large enough to respect all these lower bounds. 
Also, the sector refinement in the SOS technique has to be iteratively
carried out until Gabriel's condition is satisfied in a spherical boundary
layer. With the exception of this iterative procedure, all steps
of the algorithm are straightforward.

\section{Discussion}
This paper is mainly of theoretical interest. Implementations of existing algorithms have generated meshes that are better than what is guaranteed by my algorithm (and prior algorithms) because the bounds are derived for pathologically pessimistic circumstances. An efficient practical implementation, however, deserves a paper of its own. This aspect should be the next focus of research. Shewchuk and Si~\cite{SS14} identified circumstances where they were unable to find an implementation (for truly Delaunay mesh generation) that was successfully able to generate a mesh for a domain with multiple facets at small dihedral angles. A practical implementation may involve an octree data structure and traversal algorithm to efficiently compute the LFS for refining a facet using the advancing-front algorithm (because of spatial locality). Such data structures and traversal algorithms may be of independent interest. 

In this paper, I have not explored avenues for avoiding the construction of sliver elements in the mesh. Li and Teng~\cite{LT01} developed a technique to avoid slivers by carefully placing Steiner vertices such that a minimum bound on the dihedral angle in a Delaunay mesh exists. In the process of careful placement, they sacrifice some of the guarantees on radius-edge bounds. It should be possible to incorporate their idea into this paper, but their bounds are so small that they did not bother computing them explicitly. Cheng et al.~\cite{CD02,CDR05} developed a weighted Delaunay refinement algorithm to avoid slivers. This technique may be incorporated into my algorithm, but the bounds are too small to be meaningful. 

Obtaining a meaningful bound on the dihedral angle may not be far-fetched using my advancing-front algorithm. Labelle~\cite{L06} developed a lattice-based refinement technique, where Steiner vertices are placed only on points on a lattice. His algorithm works only on a periodic set of points, which are far away from the domain boundary. Shewchuk~\cite{She12}  suggested that an advancing-front Delaunay refinement algorithm with Labelle's technique may provide a meaningful bound on dihedral angles in a mesh. This direction is the most promising direction for future research that is of theoretical and practical interest. 

\pagebreak
\bibliographystyle{abbrv}
\bibliography{myrefs}

\end{document}